\documentclass[manuscript]{aastex}

\usepackage{natbib}
\usepackage{epsfig}
\usepackage{epstopdf}
\usepackage{morefloats}

\newcommand{\teff}{T_{\rm eff}}
\newcommand{\mj}{\rm M_{J}}

\newcommand{\ik}{{\it Kepler~}}
\newcommand{\ikt}{{\it Kepler}}

\newcommand{\sen}{$\mathcal{S}$~}
\newcommand{\sent}{$\mathcal{S}$}

\newcommand{\gs}{$\mathcal{GS}$~}
\newcommand{\gst}{$\mathcal{GS}$}

\newcommand{\gsr}{$\mathcal{GS}/\sigma_\mathcal{GS}$~}
\newcommand{\gsrt}{$\mathcal{GS}/\sigma_\mathcal{GS}$}

\newcommand{\sgst}{$\sigma_\mathcal{GS}$}

\newcommand{\lc}{light curve~}
\newcommand{\lct}{light curve}

\newcommand{\Lc}{Light curve~}

\newcommand{\pv}{{\it p}-value~}
\newcommand{\pvt}{{\it p}-value}

\newcommand{\figr}[1]{Figure~\ref{fig:#1}}
\newcommand{\secr}[1]{Section~\ref{sec:#1}}
\newcommand{\eqr}[1]{Equation~(\ref{eq:#1})}

\newcommand{\tabr}[1]{\mbox{Table~\ref{tab:#1}}}

\shorttitle{Spot-induced TTV}
\shortauthors{Holczer et al.}

\begin{document}

\title{Time variation of \ik transits induced by stellar spots --- \\ 
a way to distinguish between prograde and retrograde motion.\\
II. Application to KOIs}


\author{
Tomer Holczer\altaffilmark{1},
Avi Shporer\altaffilmark{2, 3, 10},
Tsevi Mazeh\altaffilmark{1, 4},
Dan Fabrycky\altaffilmark{5},
Gil Nachmani\altaffilmark{1},
Amy McQuillan\altaffilmark{1},
Roberto Sanchis-Ojeda\altaffilmark{6, 10},
Jerome A.~Orosz\altaffilmark{7},
William F.~Welsh\altaffilmark{7},
Eric B.~Ford\altaffilmark{8,9},
and
Daniel Jontof-Hutter\altaffilmark{8}
\\
}

\altaffiltext{1}{ School of Physics and Astronomy, Raymond and
Beverly Sackler Faculty of Exact Sciences, Tel Aviv University,
Tel Aviv 69978, Israel}

\altaffiltext{2}{Division of Geological and Planetary Sciences,
California Institute of Technology, Pasadena, CA 91125, USA}

\altaffiltext{3}{Jet Propulsion Laboratory, California Institute
of Technology, 4800 Oak Grove Drive, Pasadena, CA 91109, USA}

\altaffiltext{4}{Kavli Institute for Theoretical Physics, 
University of California, Santa Barbara}

\altaffiltext{5}{Department of Astronomy and Astrophysics, University of
Chicago, 5640 South Ellis Avenue, Chicago, IL 60637, USA}

\altaffiltext{6}{Department of Astronomy, University of California Berkeley,
Berkeley, CA 94720, USA}

\altaffiltext{7}{Astronomy Department, San Diego State
University, San Diego, CA 92182, USA}

\altaffiltext{8}{Department of Astronomy and Astrophysics, The Pennsylvania
State University, 525 Davey Laboratory, University Park,
PA 16802, USA}

\altaffiltext{9}{Center for Exoplanets and Habitable Worlds, The Pennsylvania
State University, University Park, PA 16802, USA}

\altaffiltext{10}{Sagan Fellow}

\begin{abstract}

Mazeh, Holczer, and Shporer (2015) have presented an approach that can, in principle, use the derived transit timing variation (TTV) of some transiting planets observed by the \ik mission to distinguish between prograde and retrograde motion of their orbits with respect to their parent stars' rotation. The approach utilizes TTVs induced by spot-crossing events that occur when the planet moves across a spot on the stellar surface, looking for a correlation between the derived TTVs and the stellar brightness derivatives at the corresponding transits. This can work even in data that cannot temporally resolve the spot-crossing events themselves. Here we apply this approach to the \ik KOIs, identifying nine systems where the photometric spot modulation is large enough and the transit timing accurate enough to allow detection of a TTV-brightness-derivatives correlation. Of those systems five show highly significant prograde motion (Kepler-17b, Kepler-71b, KOI-883.01, KOI-895.01, and KOI-1074.01), while no system displays retrograde motion, consistent with the suggestion that planets orbiting cool stars have prograde motion. All five systems have impact parameter $0.2\lesssim b\lesssim0.5$, and all systems within that impact parameter range show significant correlation, except HAT-P-11b where the lack of a correlation follows its large stellar obliquity. Our search suffers from an observational bias against
detection of high impact parameter cases, and the detected sample is extremely small. Nevertheless, our findings may suggest that stellar spots, or at least the larger ones, tend to be located at a low stellar latitude, but not along the stellar equator, similar to the Sun.

\end{abstract}

\section{Introduction}
\label{intro}

Observed characteristics of exoplanet systems may contain
clues to their formation process and orbital evolution history.
One of the promising clues is stellar obliquity, the angle
between the stellar spin and planetary orbital angular momentum,
also called the spin-orbit angle. Theoretical investigations have
identified processes that are expected to result in well aligned
systems \citep[e.g.,][]{cresswell07, nagasawa08, winn10a, dawson14}, and
others expected to produce misaligned orbits
\citep[e.g.,][]{fabrycky07, naoz11, batygin12}. Those theories
are now being put to the test using the increasing sample of host
stars with measured obliquity \citep[e.g.,][]{triaud10,
albrecht12, albrecht13}.

The growing sample has led to the detection of several
possible observational trends. \cite{fabrycky09} have identified
early on that the sample of obliquity measurements is better
explained as being composed of two different underlying
distributions, one well aligned and the second isotropic, than a
single distribution. Later, \cite{winn10a} suggested that cool
stars, below $\teff \approx$ 6,250 K, reside in well aligned
systems and that the obliquity of hotter stars cover a wider range. This
was supported by a study of a larger sample by \cite{albrecht12}
and a statistical study by \cite{mazeh15b}. Another trend,
identified by \cite{hebrard11}, shows that planets in a
retrograde orbit are less massive than a mass cut-off of about
$3.5\ \mj$, while planets above that threshold tend to be in
prograde orbits. This suggests that different
orbital evolution processes are at play above and below that mass
threshold. More recently, \cite{morton14} showed that stars
hosting a multi planet system (as detected by \ikt) tend to be
more well aligned than stars hosting a single planet system.

The need for a large sample of systems with a measured host star
obliquity has led to the development of several methods for
measuring it. Those include utilizing the Rossiter-McLaughlin
(RM) effect \citep[e.g.,][]{gaudi07}, asteroseismology
\citep[e.g.,][]{gizon03, chaplin13, vaneylen14, lund14}, stellar rotation
\citep[e.g.,][]{schlaufman10, hirano12, hirano14, morton14,
mazeh15b}, stellar gravity darkening \citep{barnes09, szabo11,
barnes11}, the beaming effect (Photometric RM---\citealt{shporer12, groot12}), 
and stellar activity in the form of spots \citep[e.g.,][]{nutzman11, sanchis11a, sanchis11b,
desert11, deming11, sanchis12, sanchis13}.

The methods above are either statistical in nature, so no
conclusions can be made about the configuration of a specific
system, or require additional data---spectroscopic and/or \ik
short cadence light curves.
Mazeh, Holczer, and Shporer 2015 (Hereafter Paper I) presented a
new method to distinguish between prograde and retrograde motion
of transiting systems for photometrically
active host stars. The method utilizes spot-crossing events, where the
transiting object, a planet in our case, moves across a dark spot
on the surface of the host star. Unlike the approach of
\cite{nutzman11} and \citet[][see also
\citealt{sanchis12}]{sanchis11a}, which usually requires \ik
short cadence data, this method can use \ik long cadence data, available for all objects observed by \ikt.

Briefly, the method presented in Paper I is based on measuring
two parameters for each transit event. The first is the {\it
local stellar brightness temporal derivative} at transit time, or simply the local
slope, a parameter that captures the host star's rotation phase
as the spot is rotating towards or away from the center of the
stellar disc. The second parameter is the {\it transit timing
variation} (TTV), assumed to be induced by the spot-crossing
event \citep[e.g.,][]{oshagh13}. Spot crossings
distort the transit light curve shape, and although with long
cadence data they usually cannot be fully resolved, they result
in a small mid-transit time shift when fitted with a simple
transit light curve model that does not account for the
spot-crossing event. The TTV reflects the phase within the
transit where the spot-crossing occurred. As shown in detail in
Paper I, a negative (positive) correlation between the TTV and
the local slope indicates a prograde (retrograde) motion, under
some simplistic assumptions.

Here we present the application of the method presented in Paper I
to \ik KOIs. In Section~2 we describe our analysis and the way we identify objects with
statistically significant correlation between the TTVs and local
slopes. In Section~3 we discuss the transit impact parameter of the systems expected to be sensitive to TTV due to spot-crossing. In Section~4 we present the six systems for which we
detect a statistically significant correlation, while in Section~5 we discuss the systems that are expected to show significant TTV due to spot-crossing although no significant correlation is
detected. Section~6 presents a brief discussion and a summary.

\section{Identification of KOIs with significant Correlation}
\label{sec:interesting}

As mentioned above, our method uses two parameters
measured for each transit event---the TTV and the local
slope of the stellar brightness at the time of the transit.
\secr{sample} briefly describes the catalog of transit timing
measurements we use here, and \secr{slope} describes our
measurement of the local photometric slopes. In
\secr{correlation} we derive the correlation between the two sets
of parameters for each KOI, and point to the significant detections. In 
\secr{sensitivity} we discuss the sensitivity of the different systems for detecting the correlation and compare the detected systems with their sensitivity. 

\subsection{Transit timings}
\label{sec:sample}

We used the transit timing catalog obtained by \cite{holczer15}, publicly available at:
ftp://wise-ftp.tau.ac.il/pub/tauttv/TTV/ver$\_$112.
The catalog includes all the KOIs listed in the NASA Exoplanet Archive\footnote{http://exoplanetarchive.ipac.caltech.edu/}, except KOIs

\begin{itemize}
\item labeled as false positives,
\item with orbital periods longer than 300 days,
\item with transit depth larger than 10 \%,
\item with phase folded transit signal-to-noise ratio (SNR) below 7.1, 
\item suspected to be binaries because of a difference in transit depth between odd and even transits,
\item with folded light curve that did not display a transit with high enough significance;   the \pv of the  $\mathcal{F}$-test between the transit model and the no-variability model (flat line) did not yield a level of significance higher than 10$^{-4}$.
\end{itemize}
After removing all these KOIs, we are left with 2,600 systems.

The full list  includes a total of 295,373
individual transits. Of those, 71,240 were rejected 
based on (1) an $\mathcal{F}$-test comparing the individual transit model
with a flat light curve (no variability), (2) a strong deviation
of the transit depth, duration, or timing residual (after
subtracting a linear ephemeris) from the sample mean, or (3) the
transit is too close to, or overlapping with, a transit of
another planet in the same system (see \cite{holczer15} for details). This left 224,133 entries.

\subsection{The local photometric slope}
\label{sec:slope}

We derived the local photometric slope at each transit 
by fitting a polynomial to the stellar light curve, extending
four transit durations centered on the transit, while ignoring
the in-transit data. For each transit event we normalized the out-of-transit data to
the median brightness of the relevant \ik quarter.

The fitting was done in two steps. 
First, we fitted independently six polynomials,
of degrees of one through six, using  the
\begin{footnotesize}MATLAB/REGRESSION\end{footnotesize} function that exercises the
linear least squares approach. The initial
fit was followed by a final fit, after outliers beyond four times
the scatter\footnote{Throughout this work we define the scatter as 1.4826 times the median absolute
deviation (MAD), which equals the standard deviation for a Gaussian
distribution.} were ignored.
In the second step we chose the best polynomial fit for each
transit event by performing $\mathcal{F}$-tests between all pairs
of polynomial fits and calculating their \pvt s. We chose the
best fit to be the one with the highest degree for which the \pvt s 
of all the $\mathcal{F}$-test pairs with polynomial fits of
lower degrees were lower than 10$^{-3}$. 
To obtain the local brightness slope during the transit we 
calculated the derivative of the best-fit polynomial at
mid-transit time. The derivative uncertainty was propagated from the errors of the polynomial coefficients. 

Next, we examined the sample of local slopes for each KOI and
removed outlier slopes. A slope was identified as an outlier if
it deviated from the sample mean by more than five times the
sample scatter plus three times the uncertainties median. Among
the 224,133 measured local slopes we identified 265 (= 0.12 \%)
as outliers, leaving 223,868 to be used in the following
analysis.

\tabr{slope} lists all transit events we analyzed, including the
fitted local slope, $s$, the local slope uncertainty, $\sigma_s$, the
degree of the best-fit polynomial, and a flag marking outlier
local slopes.

\begin{table}[!ht]
\footnotesize \caption{\Lc local slope for KOI transits}
\begin{tabular}{rrrrrrr}
\hline \hline
KOI & n\tablenotemark{a}  & $t_n$\tablenotemark{b}~~~&
$s$\tablenotemark{c}~~ &   $\sigma_s$\tablenotemark{d}~~  &
deg\tablenotemark{e} & flag\tablenotemark{f}\\
      &                                   &  [d]~~~
& [ppm/d]                          & [ppm/d]
&                                      &
\\

\hline
$ 1.01 $ & $ 0 $ & $ 55.7633 $ & $ -39 $ & $ 66 $ & $ 1 $ & $ 0 $
\\
$ 1.01 $ & $ 1 $ & $ 58.2340 $ & $ -47 $ & $ 81 $ & $ 1 $ & $ 0 $
\\
$ 1.01 $ & $ 2 $ & $ 60.7046 $ & $ 132 $ & $ 68 $ & $ 1 $ & $ 0 $
\\
$ 1.01 $ & $ 4 $ & $ 65.6458 $ & $ 19 $ & $ 70 $ & $ 1 $ & $ 0 $ \\
$ 1.01 $ & $ 5 $ & $ 68.1164 $ & $ -40 $ & $ 50 $ & $ 1 $ & $ 0 $ \\
$ 1.01 $ & $ 6 $ & $ 70.5870 $ & $ -59 $ & $ 83 $ & $ 1 $ & $ 0 $ \\
$ 1.01 $ & $ 7 $ & $ 73.0576 $ & $ -6 $ & $ 72 $ & $ 1 $ & $ 0 $
\\
$ 1.01 $ & $ 8 $ & $ 75.5282 $ & $ 14 $ & $ 75 $ & $ 1 $ & $ 0 $
\\
$ 1.01 $ & $ 9 $ & $ 77.9989 $ & $ 17 $ & $ 79 $ & $ 1 $ & $ 0 $
\\
$ 1.01 $ & $ 10 $ & $ 80.4695 $ & $ -19 $ & $ 58 $ & $ 1 $ & $ 0 $ \\
\hline
\end{tabular}
\tablenotetext{a}{Transit number.}
\tablenotetext{b}{Expected transit time of the linear ephemeris
in BJD --
2454900 following \cite{holczer15}.}
\tablenotetext{c}{Derived local slope.}
\tablenotetext{d}{Derived local slope uncertainty.}
\tablenotetext{e}{Polynomial degree of the best fit chosen.}
\tablenotetext{f}{Transit flag: 1 = transit identified as an
outlier. 0 = not an outlier.}\\
(This table is available in its entirety in a machine-readable
form in ftp://wise-ftp.tau.ac.il/pub/tauttv/Slope. A portion of
the
table is shown here for guidance regarding its form and
content.)
\label{tab:slope}
\end{table}
%

\subsection{Searching for slope-TTV (anti)correlation}
\label{sec:correlation}

We now turn to identify KOIs with a statistically significant
correlation between the TTV and the local photometric slope.
We chose to derive the correlation by fitting a linear function
to the TTV vs.~the local slope (using the
\begin{footnotesize}MATLAB/REGRESSION\end{footnotesize}
function), and then identify the KOIs for which the slope of that linear
function strongly deviates from zero.

We referred to the fitted linear function's slope as the {\it global
slope}, \gst, with an uncertainty \sgst. We used the absolute value of the ratio of the two, \gsrt, as a proxy to the correlation's statistical significance. We performed this fit
only for KOIs with 20 or more measured transits, 
to avoid the effects of small number statistics. Out
of 2,600 KOIs, 1,858 had 20 or more measurements. To assign a
false alarm probability (FAP) to each correlation, we performed
$10^7$ bootstrap tests, randomly permuting the sets of local
slopes. We defined the \pv of the derived global slope as the
fraction of permutations that yielded \gsr larger in its absolute
value than that of the real data set.

\tabr{statistic} lists the statistical parameters for the
KOIs in the sample. It also includes the orbital period, transit depth, and the
TTV uncertainties, taken from
\cite{holczer15}\footnote{ftp://wise-ftp.tau.ac.il/pub/tauttv/TTV
/ver$\_$112}, and the number of transits for which both the local
slope and the TTV are not rejected as outliers. The fitted \gst, \sgst, and the
corresponding \pv are listed for 1,858 KOIs which had 20 or more
measurements.

\begin{table}[!h]
\footnotesize \caption{Statistical parameters of the TTVs and the
stellar rotation of the KOIs}
\begin{tabular}{|r|rr|rrr|rrr|rrrr|}
\hline \hline

KOI &
Period\tablenotemark{a} &
Depth\tablenotemark{b}&
Period\tablenotemark{c}      &
Amp.\tablenotemark{d}  &
flag\tablenotemark{e} &
$\mathcal{GS}$\tablenotemark{f}   ~&
$\sigma_\mathcal{GS}$\tablenotemark{g} ~ &
p-($\mathcal{GS}/\sigma_\mathcal{GS}$)\tablenotemark{h} &
Expected\tablenotemark{i} &
$\sigma_{TTV}$\tablenotemark{j}&
N\tablenotemark{k} &
$\mathcal{S}\tablenotemark{l}$ \\
       &        Orb.    \    & & Rot.    \    & Rot.              &
& & & & TTV~~ & & & \\
       &
       [d]~ ~  &        [ppm]&        [d]   ~~   &
       [ppm]                         &        &        [min d]&
       [min d]             &        [log] ~ &        [min] ~ &        [min] &   &\\
\hline
$ 1.01 $ & $ 2.47 $ & $ 14210 $ & \nodata & \nodata & $ 0 $ &
$ -46 $ & $ 61 $ & $ 0.0 $ & \nodata & $ 0.08 $ & $ 428 $ &
\nodata \\
$ 2.01 $ & $ 2.20 $ & $ 6694 $ & \nodata & \nodata & $ 0 $ &
$ -32 $ & $ 76 $ & $ -0.1 $ & \nodata & $ 0.23 $ & $ 599 $ &
\nodata \\
$ 3.01 $ & $ 4.89 $ & $ 4361 $ & $ 29.47 $ & $ 5875 $ & $ 1 $ & $
26 $ & $ 19 $ & $ -0.7 $ & $ 2.97 $ & $ 0.24 $ & $ 214 $ & $
178.8 $ \\
$ 5.01 $ & $ 4.78 $ & $ 980 $ & \nodata & \nodata & $ 0 $ &
$ -300 $ & $ 1000 $ & $ 0.0 $ & \nodata & $ 1.65 $ & $ 278 $ &
\nodata \\
$ 7.01 $ & $ 3.21 $ & $ 736 $ & \nodata & \nodata & $ 0 $ & $
1100 $ & $ 1900 $ & $ -0.3 $ & \nodata & $ 3.53 $ & $ 325 $ &
\nodata \\
$ 10.01 $ & $ 3.52 $ & $ 9370 $ & \nodata & \nodata & $ 0 $ & $
130 $ & $ 110 $ & $ -0.6 $ & \nodata & $ 0.60 $ & $ 380 $ &
\nodata \\
$ 12.01 $ & $ 17.86 $ & $ 9228 $ & $ 1.25 $ & $ 390 $ & $ 1 $ & $
70 $ & $ 150 $ & $ -0.2 $ & $ 0.45 $ & $ 0.37 $ & $ 71 $ & $ 10.3
$ \\
$ 13.01 $ & $ 1.76 $ & $ 4602 $ & \nodata & \nodata & $ 0 $ & $
196 $ & $ 62 $ & $ -4.7 $ & \nodata & $ 0.13 $ & $ 735 $ &
\nodata \\
$ 17.01 $ & $ 3.23 $ & $ 10811 $ & \nodata & \nodata & $ 0 $ & $
50 $ & $ 120 $ & $ -0.7 $ & \nodata & $ 0.33 $ & $ 334 $ &
\nodata \\
$ 18.01 $ & $ 3.55 $ & $ 7454 $ & \nodata & \nodata & $ 0 $ & $
1060 $ & $ 190 $ & $ -7.0 $ & \nodata & $ 0.57 $ & $ 376 $ &
\nodata \\
\hline
\end{tabular}
\tablenotetext{a}{Orbital period.}
\tablenotetext{b}{ Transit depth.}
\tablenotetext{c}{Rotation period.}
\tablenotetext{d}{Rotation semi amplitude (see
\citet{mcquillan13,mcquillan14} for details).}
\tablenotetext{e}{Rotation Flag, based on the findings of
\cite{mcquillan13, mcquillan14}. 0: KOI was analyzed and no
period was identified, 1: A rotation period was detected, 2: KOI was not analyzed.
}
\tablenotetext{f}{Global slope. The slope of the linear fit to the TTV as a
function of the local slope.}
\tablenotetext{g}{Global slope uncertainty.}
\tablenotetext{h}{{\rm Log} \pv of
$\mathcal{GS}/\sigma_\mathcal{GS}$.}
\tablenotetext{i}{The maximum expected TTV.}
\tablenotetext{j}{The TTV uncertainties median.}
\tablenotetext{k}{Number of measurements.}
\tablenotetext{l}{TTV Sensitivity.}
\\
(This table is available in its entirety in a machine-readable
form in ftp://wise-ftp.tau.ac.il/pub/tauttv/Slope. A portion is shown
here for guidance regarding its form and content.)
\label{tab:statistic}
\end{table}

Our search could be applied only to active stars with large enough spot-induced stellar modulation.  
We expected those stars to be identified by \citet{mcquillan13, mcquillan14}, who searched the \ik light curves for stellar rotation. Therefore while our analysis (as described above)
was applied to all 2,600 KOIs, we have applied a more detailed analysis (as described below) to the subsample of 862 KOIs whose host star
rotation period and photometric activity amplitude were measured
by \citet{mcquillan13, mcquillan14}. For KOIs in that subsample we have calculated the maximum expected TTV due to spot-crossing,
max$\{TTV_{sc}\}$, by Equation~(13) of Paper I, using the
published stellar rotation period and amplitude, all listed in
\tabr{statistic}. Of those 862 KOIs, only 726 had at least 20
measurements of the TTV and the local slope.

\subsection{Detection sensitivity of the slope-TTV (anti)correlation}
\label{sec:sensitivity}

We now turn to discuss the sensitivity of each KOI's data to
the detection of the  TTV--local-slope correlation,  
assuming the time variation is induced by
spot-crossing events. The detection depends on the magnitude of
the induced TTVs, in terms of the TTV uncertainties, and on the
square root of the number of measurements (i.e., transits).
Therefore we define a new quantity, \sent, to be:
\begin{equation}
\label{eq:sen}
\mathcal{S} = \frac{{\rm max}\{TTV_{sc}\}}{\sigma_{TTV}}
\sqrt{N}\,,
\end{equation}
where $\sigma_{TTV}$ is the median TTV uncertainty and $N$ the number of transit timing measurements.
We note that we can derive \sen only for the 862 systems for
which the photometric modulation of the stellar rotation was
obtained. For those systems, all four quantities in
\eqr{sen} are listed in \tabr{statistic}.

\begin{figure}[h!]
\centering
{\includegraphics[width=15cm]{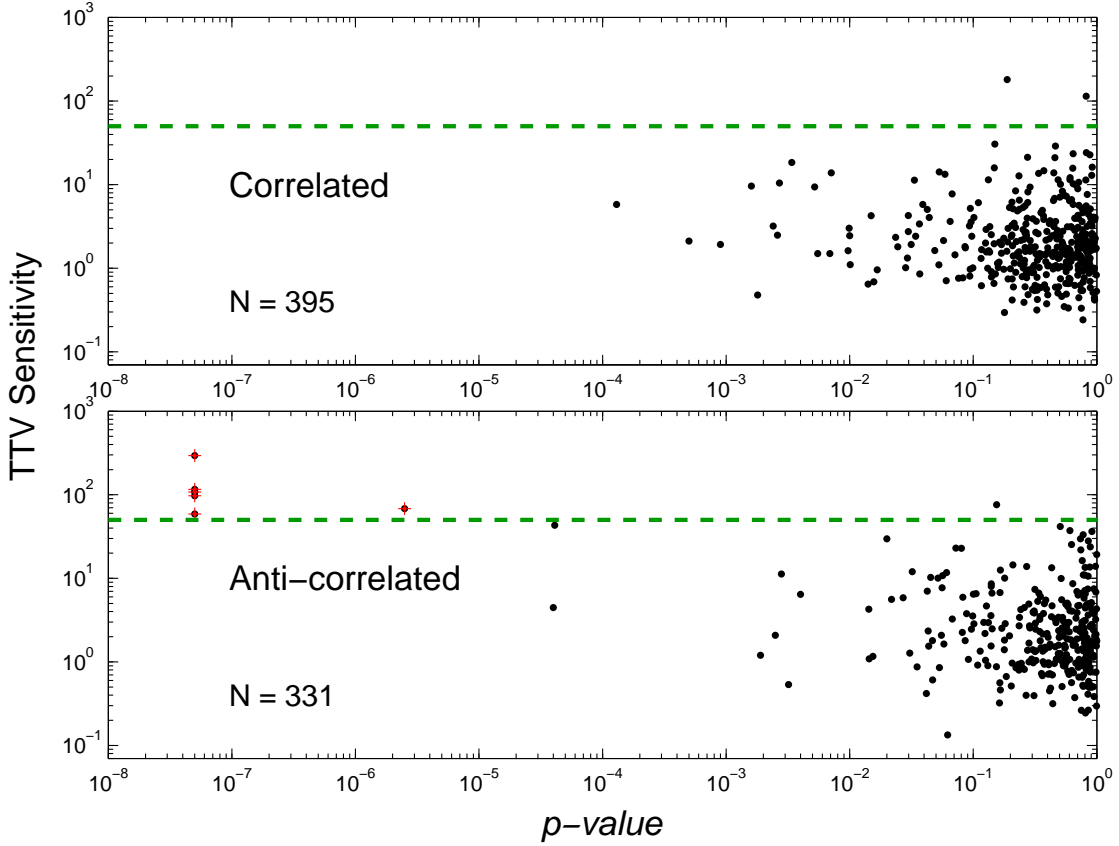}}
\caption{Sensitivity to the correlation of TTV induced by spot-crossing with the brightness slope, \sent,
vs.~the derived statistical significance of the correlation for the sample of 726 KOIs with detected stellar rotation and at least 20 transits. The
upper panel shows KOIs for which we obtained a positive slope and
the lower panel shows KOIs with a negative slope.
The red crosses are the KOIs with \pvt s lower than 10$^{-5}$
and the green dashed line marks an expected TTV sensitivity of
\sen = 50. The latter is a somewhat arbitrary value, used as a
threshold to mark KOIs with a strong TTV sensitivity. 
}
\label{fig:expectedTTV}
\end{figure}

In \figr{expectedTTV} we plot \sen against the derived \pv of the derived slope in
logarithmic scale. Only the 726 KOIs with measured
stellar rotation and more than 20 transits are plotted, as only those
allow to calculate \sent. The upper (lower) panel shows KOIs with
positive (negative) correlations. 
The dashed line in both panels
is at \sent=50, chosen somewhat arbitrarily as a threshold to
identify KOIs with a strong sensitivity to TTV induced by
spot-crossing. 

%

In order to single out the systems with significant correlation detection we
picked out the KOIs that have a \pv lower than
10$^{-5}$. We found six such KOIs with negative \gst, suggesting
a prograde configuration, while not a single system
with positive \gst. In fact, we did not find any system with a
positive \gs and a \pv below 10$^{-4}$. 
These six KOIs with significant correlation are marked in the figure by red crosses, and are discussed in detail in \secr{KOIs_significant}.

One would expect to find systems showing a strong correlation (or
anti-correlation) due to spot-crossings in the upper left corner,
with high sensitivity and low \pvt. Indeed, we find the six KOIs
with significant correlation in that corner in the lower panel, while
the same corner in the upper panel is empty. 
For five out of these systems {\it all} our $10^7$ random permutations
gave a \gsr smaller (in absolute value) than that of the real
data, so somewhat arbitrarily we assigned them a \pv of
5$\cdot$10$^{-8}$ in Figure~\ref{fig:expectedTTV}.  
We also find three
other KOIs with high sensitivity (above the dashed line in both panels of
\figr{expectedTTV}) but with no significant detection.
The nine systems and their parameters  are listed in \tabr{significant_TTV}.
This includes the stellar effective temperature
$T_{\rm eff}$, taken from the NASA
Exoplanet Archive on December 1st, 2014, and the impact parameter $b$ of the planetary orbit, derived here as described in \secr{KOIs_impact}.

\begin{table}[!h]
\scriptsize \caption{Statistical parameters of the KOIs with high
TTV sensitivity}
\begin{tabular}{|r|rr|rr|rrr|rcrr|cc|}
\hline
KOI \ \ &
Period\tablenotemark{a}&
Depth\tablenotemark{b}&
Period\tablenotemark{c}   &
Amp.\tablenotemark{d}  &
$\mathcal{GS}$\tablenotemark{e}  \ \ &
$\sigma_\mathcal{GS}$\tablenotemark{f} \  &
p-($\mathcal{GS}/\sigma_\mathcal{GS}$)\tablenotemark{g} &
Expected\tablenotemark{h} &
$\sigma_{TTV}$\tablenotemark{i}&
N\tablenotemark{j} \ &
$\mathcal{S}$\tablenotemark{k}&
$T_{\rm eff}$\tablenotemark{l} &
$b$\tablenotemark{m} \\
&   Orb. \    & & Rot.  \     & & Rot. \ $\!\!\!$   & & &
TTV \ \ & & & & & \\
& [d]  \ \ & [ppm] & [d] \ \ \       & [ppm]  & [min d] & [min d]    &
[log] \ \ & [min] \ \ & [min] &
& & [K] &\\
\hline
$ 3.01 $ & $ 4.89 $ & $ 4361 $ & $ 29.47 $ & $ 5875 $ & $ 26 $ & $ 19 $ & $ -0.7 $ & $ 2.97 $ & $ 0.24 $ & $ 214 $ & $ 178.8 $ & $ 4777 $ & $ 0.28 $ \\ 
$ 63.01 $ & $ 9.43 $ & $ 4037 $ & $ 5.41 $ & $ 8910 $ & $ 2 $ & $ 10 $ & $ -0.1 $ & $ 4.27 $ & $ 0.44 $ & $ 139 $ & $ 114.9 $ & $ 5650 $ & $ 0.73 $ \\ 
$ 203.01 $ & $ 1.49 $ & $ 21575 $ & $ 12.16 $ & $ 12865 $ & $ -33
$ & $ 3 $ & $< -7 $ & $ 3.20 $ & $ 0.29 $ & $ 710 $ & $ 292.1 $ &
$ 5624 $ & $ 0.24 $ \\
$ 217.01 $ & $ 3.91 $ & $ 22620 $ & $ 19.77 $ & $ 4970 $ & $ -140
$ & $ 20 $ & $< -7 $ & $ 1.46 $ & $ 0.45 $ & $ 331 $ & $ 58.4 $ &
$ 5543 $ & $ 0.26 $ \\
$ 254.01 $ & $ 2.46 $ & $ 40367 $ & $ 15.81 $ & $ 10930 $ & $ -9 $ & $ 6 $ & $ -0.8 $ & $ 1.57 $ & $ 0.48 $ & $ 540 $ & $ 75.4 $ & $ 3820 $ & $ 0.57 $ \\ 
$ 883.01 $ & $ 2.69 $ & $ 39186 $ & $ 9.02 $ & $ 11340 $ & $ -38
$ & $ 4 $ & $ < -7 $ & $ 1.94 $ & $ 0.40 $ & $ 495 $ & $ 107.1 $
& $ 4809 $ & $ 0.51 $ \\
$ 895.01 $ & $ 4.41 $ & $ 13822 $ & $ 5.07 $ & $ 10575 $ & $ -39
$ & $ 5 $ & $< -7 $ & $ 5.33 $ & $ 0.95 $ & $ 302 $ & $ 97.0 $ &
$ 5600 $ & $ 0.37 $ \\
$ 1074.01 $ & $ 3.77 $ & $ 14018 $ & $ 4.04 $ & $ 8360 $ & $ -32 $ & $ 6 $ & $ -5.6 $ & $ 4.12 $ & $ 1.04 $ & $ 296 $ & $ 67.9 $ & $ 6302 $ & $ 0.36 $ \\ 
$ 1546.01 $ & $ 0.92 $ & $ 15752 $ & $ 0.91 $ & $ 7600 $ & $ -27
$ & $ 1 $ & $< -7 $ & $ 1.90 $ & $ 0.62 $ & $ 1430 $ & $ 115.6 $
& $ 5713 $ & $ 0.67 $ \\
\hline
\end{tabular}
\tablenotetext{a}{Orbital period.}
\tablenotetext{b}{Transit depth.}
\tablenotetext{c}{Rotation period.}
\tablenotetext{d}{Rotation semi amplitude (see
\citealt{mcquillan13} for details).}
\tablenotetext{e}{Global slope. The slope of the linear fit to the TTV as a
function of the local slope.}
\tablenotetext{f}{Global slope uncertainty.}
\tablenotetext{g}{{\rm Log} \pv of
$\mathcal{GS}/\sigma_\mathcal{GS}$.}
\tablenotetext{h}{The maximum expected TTV.}
\tablenotetext{i}{The TTV uncertainties median.}
\tablenotetext{j}{Number of measurements.}
\tablenotetext{k}{TTV Sensitivity.}
\tablenotetext{l}{Stellar effective temperature taken from the NASA Exoplanet Archive. Typical uncertainty is 100--200 K.}
\tablenotetext{m}{Impact parameter derived here using \ik short cadence data (see \secr{KOIs_impact} below). Typical uncertainty is 0.01--0.02. For KOI-1546.01 no short cadence data is available so we used the value from the NASA Exoplanet Archive.}
\label{tab:significant_TTV}
\end{table}

%
%
\section{The impact parameter and its impact on the correlation}
\label{sec:KOIs_impact}

As discussed and demonstrated in the simulations of Paper I, the TTV local-slope correlation can be detected only if the spot crossing events occur in varying locations on the stellar disc. This can happen only if the chords of the planet and the spot lie across the stellar disc side by side. If the latitude of the spot, $\theta$, is such that $\cos\theta$ is substantially different from the impact parameter of the transiting planet, or if the stellar obliquity is substantially different from zero (or from 180$^{\circ}$), there cannot be any observed
correlation. It is therefore of interest to examine the impact parameters of the systems with high sensitivity and see if the information about the planet motion over the stellar disc can tell us something about the location of the stellar spots. This discussion does not include KOI-1546, which was recently identified to be an eclipsing stellar binary\footnote{The true nature of this object was noted on March 2015 on the Community Follow-up Observing Program (CFOP) website, see: https://cfop.ipac.caltech.edu/edit\_target.php?id=1546}.

We decided not to use the NASA Exoplanet Archive catalog values for the impact parameters of the KOIs, as they are based on long cadence data only, whereas short cadence data are available for all our high sensitivity systems.  Furthermore, the KOI sample on the NASA Exoplanet Archive displays an impact parameter distribution with a strong increase towards $b = 0$ (at the time of writing), instead of the expected uniform distribution. We do not know the cause for this artifact, but it suggests the use of those values may be unreliable for our tests.  

Therefore, we derived the impact parameters ourselves for the eight systems, using the short cadence simple-aperture photometry data. The first step was detrending the transit light curve by finding the continuum around each transit, ignoring the photometry of
the transit itself, up to 0.6 transit durations around the expected transit center. We then fitted a more extended region, of up to four durations around the expected transit center, and fitted
six different polynomials of degrees one to six. An $\mathcal{F}$-test was performed between all of the
different pairs of polynomial fits. The best fit was chosen as the one with the highest degree,
for which the $p$-value of all the $\mathcal{F}$-tests with polynomial fits of lower degrees was lower than $10^{-3}$. The last stage of the detrending process was to add back the transit points and
divide the flux at that region by the chosen polynomial fit. 

To obtain a model for the transit we folded the light curves of the different transits
together using the best-fitting period and phase. 
We then modeled the transit using the {\sc occultnl} model \citep{mandel02}, where the position of the planet was chosen with a rectilinear model of normalized impact parameter $b$, transit duration $T_{\rm dur}$ from mid-ingress to mid-egress, a linear limb darkening law with parameter $c_2$ (and other $c_i$ of \citet{mandel02} fixed to zero), and planet-to-star radius ratio $R_p/R_\star$. The model assumed no third light for the eight systems, consistent with the small contamination ($\lesssim10\%$) reported at the MAST.  

\begin{figure}[h!]
\centering
{\includegraphics[width=15cm]{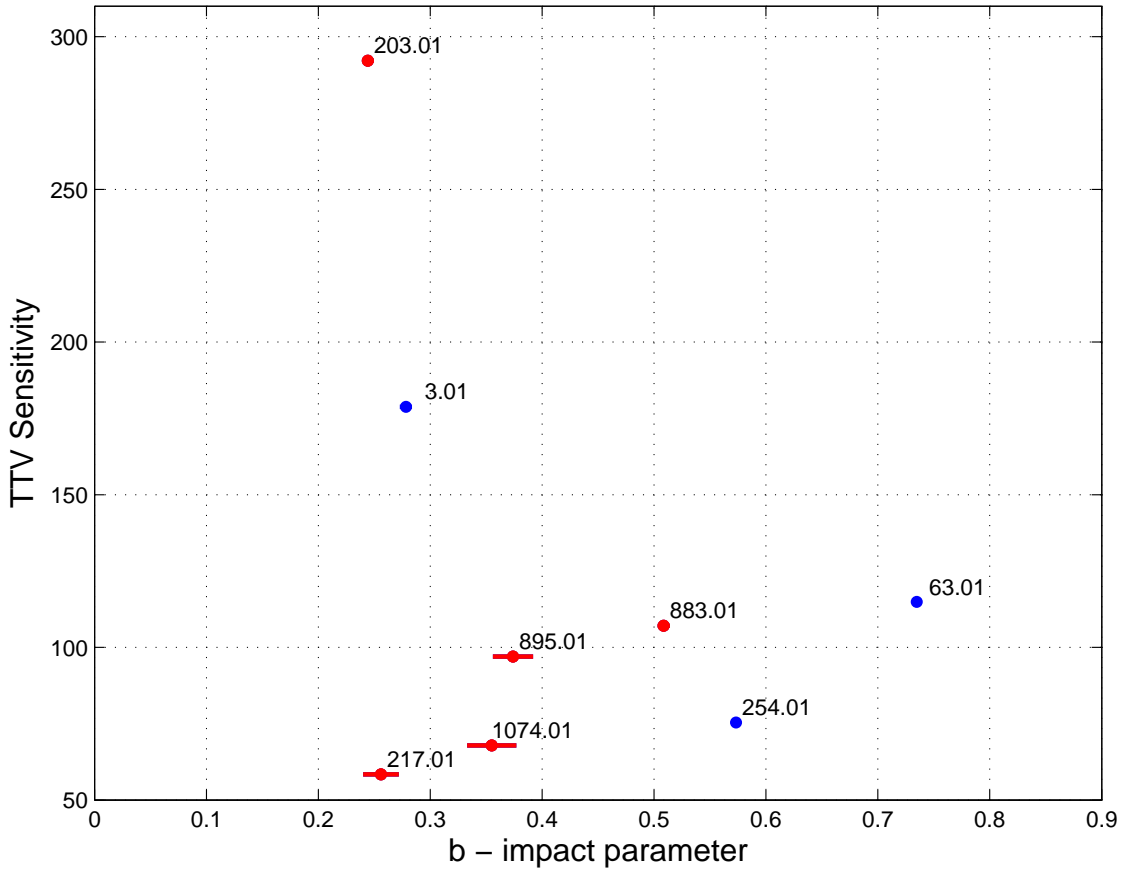}}
\caption{Sensitivity to the correlation of TTV induced by spot-crossing with the brightness slope, \sent,
vs.~the derived impact parameter of their transiting planet. Impact parameter was derived from the short cadence data, using the \cite{mandel02} model. For some systems the error bar of the impact parameter is smaller than the marker size.
The red points are the KOIs with detected correlation, with \pv lower than 10$^{-5}$, and the blue points are the systems with no significant detection of a correlation.}
\label{fig:S_vs_b}
\end{figure}

The fit was performed via the non-linear fitter {\sc IDL/mpfit} \citep{markwardt09}. 
The scatter of out-of-transit points, $\sigma$, was used to determine the uncertainty of the individual measurements, instead of relying on the reported photometric errors. Ten iterations of the fitter were run, followed by rejection of points lying more than $10\sigma$ away from the model; this procedure was repeated; and a final ten iterations of the fitter resulted in the impact parameters reported in Table~\ref{tab:significant_TTV} and plotted in Figure~\ref{fig:S_vs_b}.  
The reduced $\chi^2$ varied from $0.99$ to $1.78$ with a mean of $1.23$---these moderately exceed what white noise would have given, and they may be due to the scatter induced by the star spot crossing events.

%

Figure~\ref{fig:S_vs_b} shows that the impact parameters of the five systems displaying TTV correlated with the local slope are in the range of $0.2\lesssim b\lesssim 0.5$. Interestingly, the Sun's spots reach up to about $30^\circ$ in latitude, and their size grows smaller as they descend in latitude until reaching the equator when the Sun's activity cycle is at minimum \citep[e.g.,][]{wilson96, li00}. Therefore, to an observer viewing the Sun from the equator plane, transiting planets would cross sunspots only if they have $b \leq 0.5$, and would be difficult to detect at $b \approx 0$, just as we see in these systems. 

The only system with high sensitivity and $b \leq 0.5$ that does not show detected correlation is KOI-3.01 (= Kepler-3b = HAT-P-11b; \citealt{bakos10,winn10b, hirano11, sanchis11b}; see also \citealt{beky14}). This may be accounted for by the large stellar obliquity of KOI-3.01, as explained in more details in \secr{KOIs_non-significant}. A similar scenario might account for the lack of a correlation for KOI-63.01 (= Kepler-63b; \citealt{sanchis13}). For KOI-254.01 (= Kepler-45b; \citealt{johnson12}) there is no evidence for correlation and therefore for spot crossings, which might be explained by the relatively large impact parameter, beyond 0.5, that causes the planet to move across the stellar disk at higher latitudes than the spots. These three systems (KOI-3.01, KOI-63.01, and KOI-254.01) are reviewed in more details in \secr{KOIs_non-significant}. 

Taken together, although  extremely small, our sample suggests that the stars considered here have spots within $\approx30^{\circ}$ of their equator but not exactly along the equator, in analogy to sunspots. However, this does not exclude the existence of high latitude star spots \citep[e.g.,][]{li00} since our method is less sensitive to such systems. This is so because as the spot moves to higher latitude both the stellar photometric modulation amplitude and transit timing precision decrease.

In the next two sections we present the detailed analysis of the KOIs with high sensitivity. 
Section~\ref{sec:KOIs_significant} discusses the systems with significant correlation, and  
Section~\ref{sec:KOIs_non-significant} discusses the systems with non-significant correlation.

\section{The six KOIs with significant correlation}
\label{sec:KOIs_significant}

\begin{figure}
{\includegraphics[width=16.5cm]{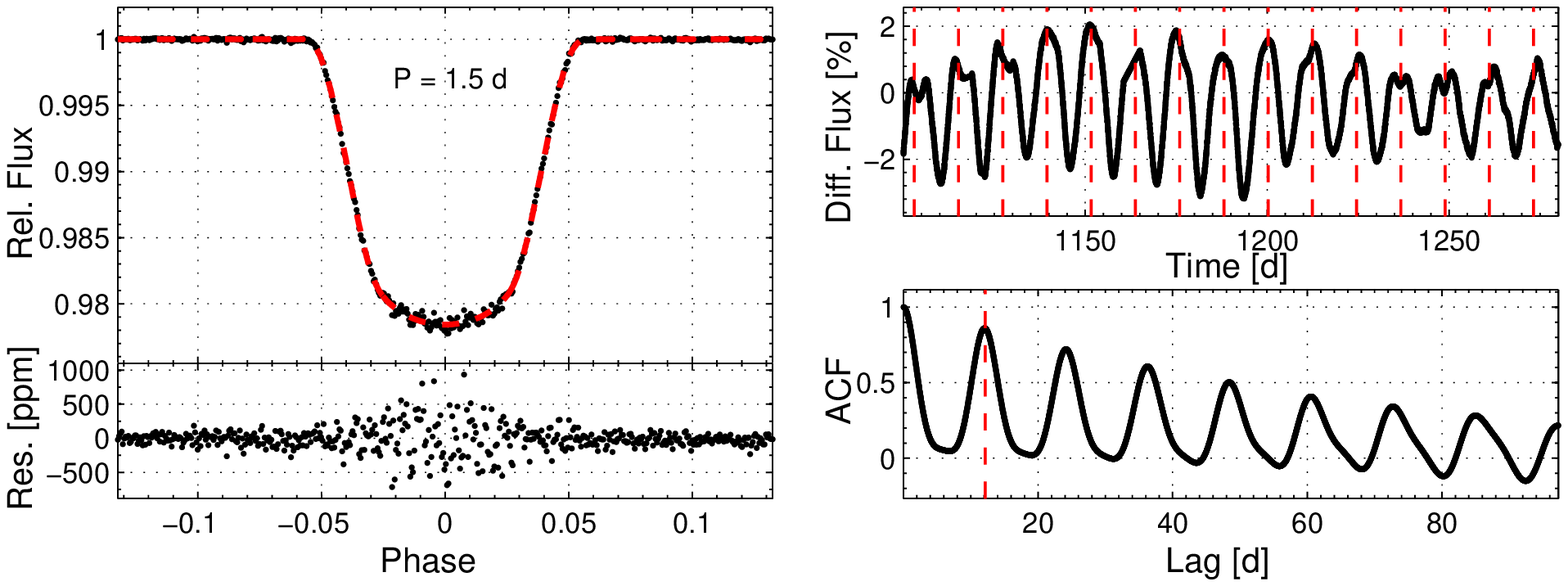}}
\caption{KOI-203.01 (= Kepler-17b): Transit light curve and
stellar
modulation and its autocorrelation.
{\it Left:} Model fit to the detrended phase folded transit \lc
(top) and its residuals (bottom, in ppm). The fitted model is
overplotted in a red dashed line.
{\it Upper right:} A segment of the \lct. The spacing between
the vertical red dashed lines equals the host star's rotation
period.
One can follow the stellar rotation modulation with the detected
period. The modulation
changes its shape and amplitude, but retains its period.
{\it Bottom right:} The Auto-Correlation Function (ACF), with a
vertical red dashed line where
the rotation period is found. One can clearly see the ACF peaks
at lags equal to integer multiples of the rotation period.
}
\label{fig:koi1}
\end{figure}

\begin{figure}
{\includegraphics[width=16.5cm]{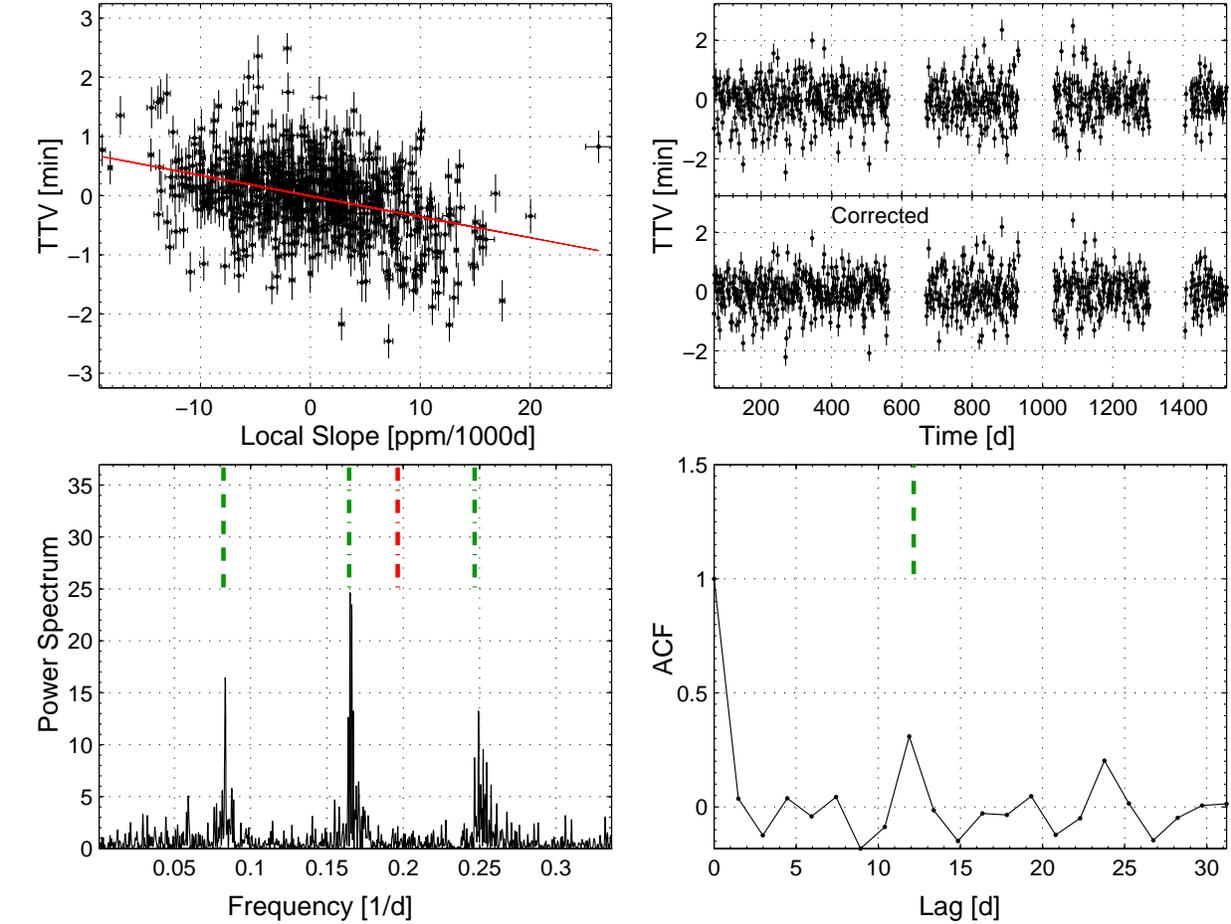}}
\caption{Analysis of the TTV of KOI-203.01 (= Kepler-17b).
{\it Upper left:} The TTV, in minutes, as a function of the
corresponding
local slope, in ppm per thousand days. The overplotted solid line
(red) is a linear fit whose slope is defined
as the global slope.
{\it Upper right:} The derived TTVs as function of time.
The upper panel shows the derived TTV, while
the lower panel displays the corrected one, after subtracting the
linear fit plotted in upper left panel.
{\it Bottom left:} The TTV power spectrum. One can
clearly see strong peaks corresponding to the stellar rotation period, marked by a green dashed line, and its first two harmonics, marked by green dotted-dashed lines. The red
dotted-dashed line marks the sampling stroboscopic frequency
which is the TTV signal induced by the sampling of the \ik long
cadence data \citep{szabo13, mazeh13}. The power spectrum is plotted up to the Nyquist frequency which in this case is half the orbital frequency.
{\it Bottom right:} The ACF of the derived TTV. The green
dashed line presents the lag corresponding to the stellar
rotation. One can notice a small peak at this lag and another one
at a lag twice the rotation period.
}
\label{fig:koi1_2}
\end{figure}

\begin{figure}
{\includegraphics[width=16.5cm]{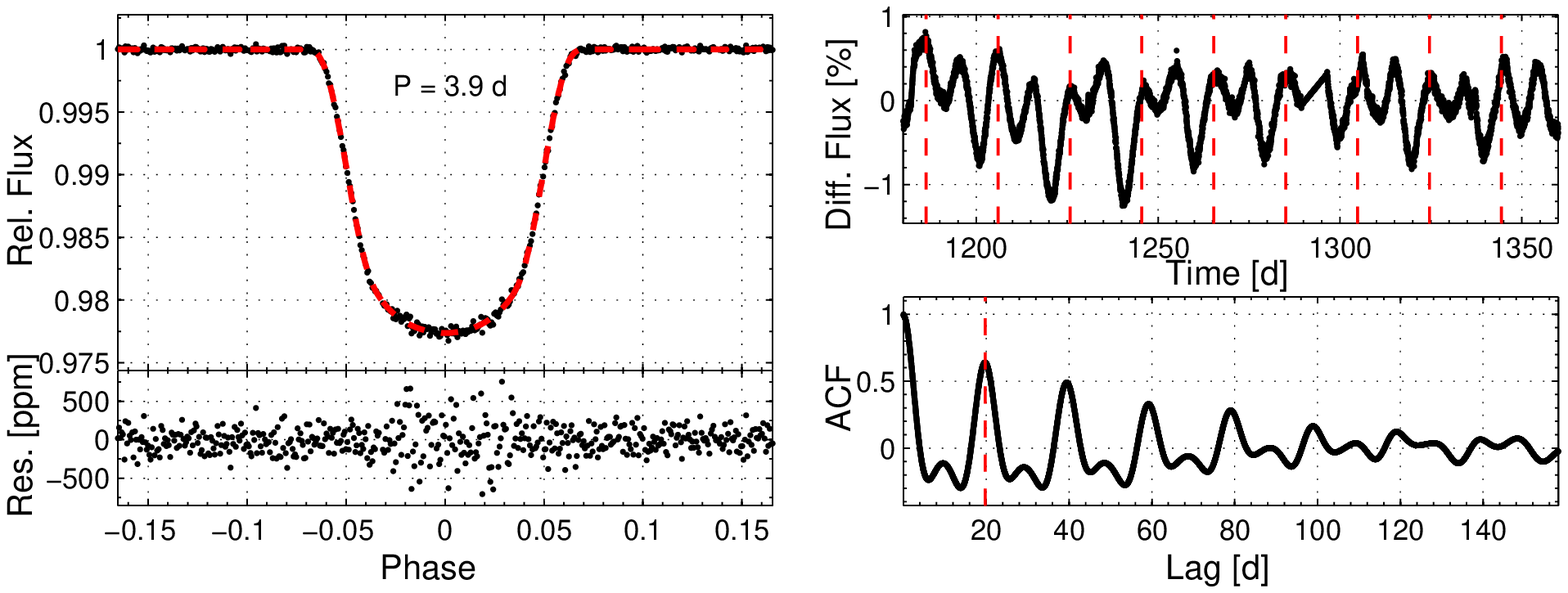}}
\caption{KOI-217.01 (= Kepler-71b) transit light curve and
stellar
modulation and its autocorrelation. See \figr{koi1} for details.}
\label{fig:koi2}
\end{figure}

\begin{figure}
{\includegraphics[width=16.5cm]{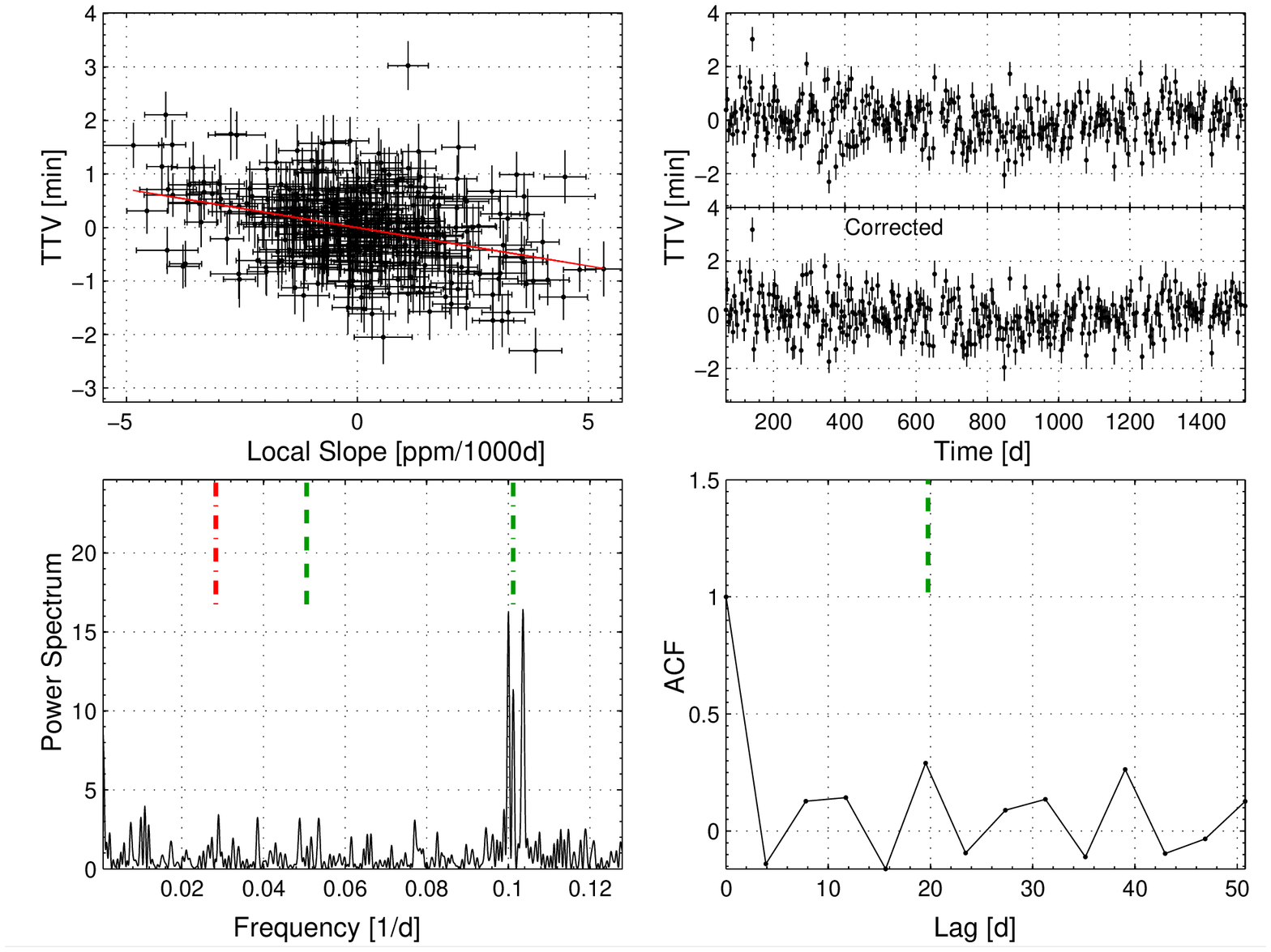}}
\caption{Analysis of the TTV of KOI-217.01 (= Kepler-71b). See
\figr{koi1_2} for details. 
The green dashed line in the bottom left panel mark the stellar rotation frequency, and its first harmonic is marked by a green dotted-dashed line.}
\label{fig:koi2_2}
\end{figure}

\begin{figure}
{\includegraphics[width=16.5cm]{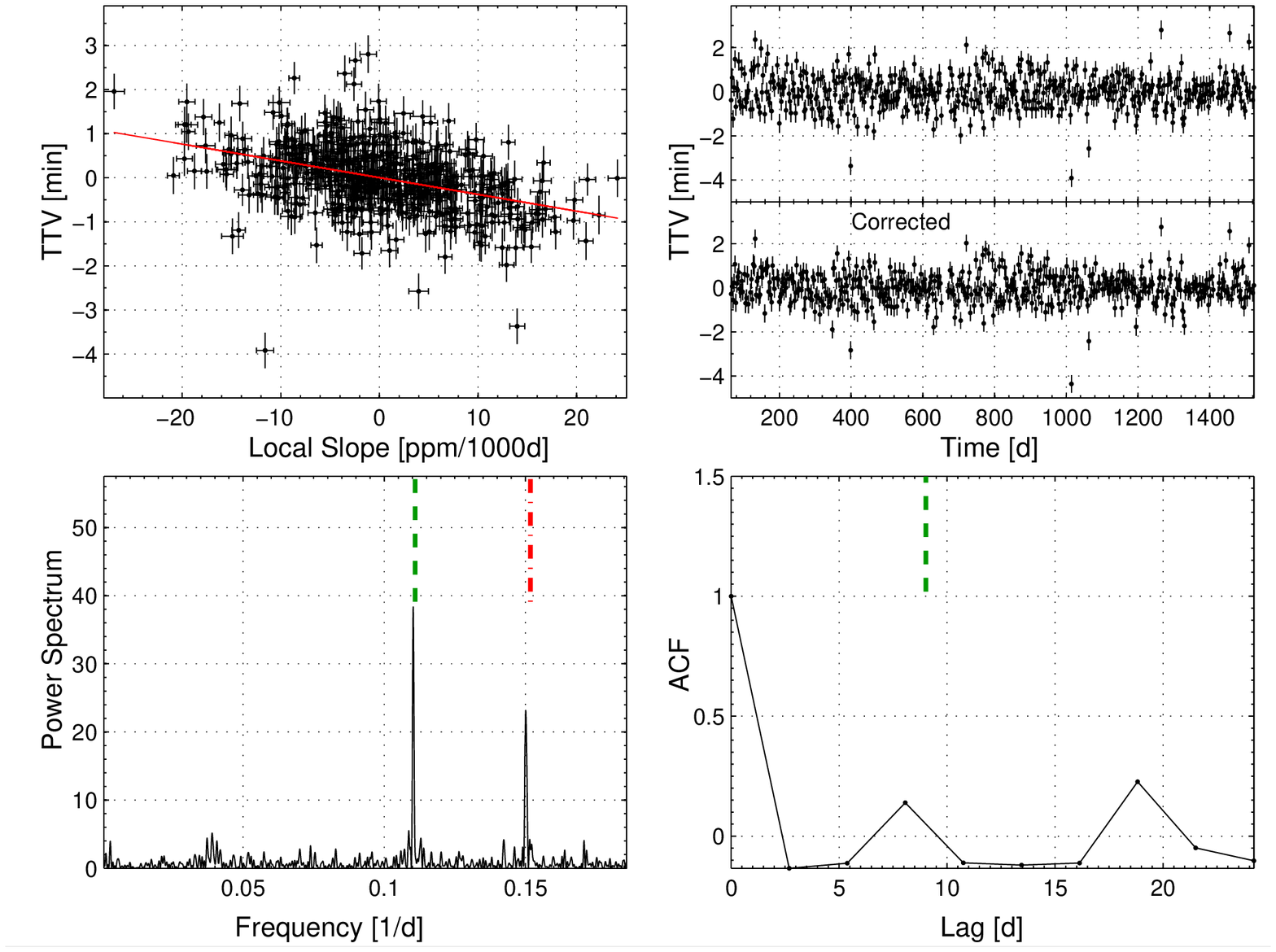}}
\caption{Analysis of the TTV of KOI-883.01. See \figr{koi1_2} for
details.
The green dashed line in the bottom left panel marks the stellar rotation frequency.
 }
\label{fig:koi3_2}
\end{figure}

\begin{figure}
{\includegraphics[width=16.5cm]{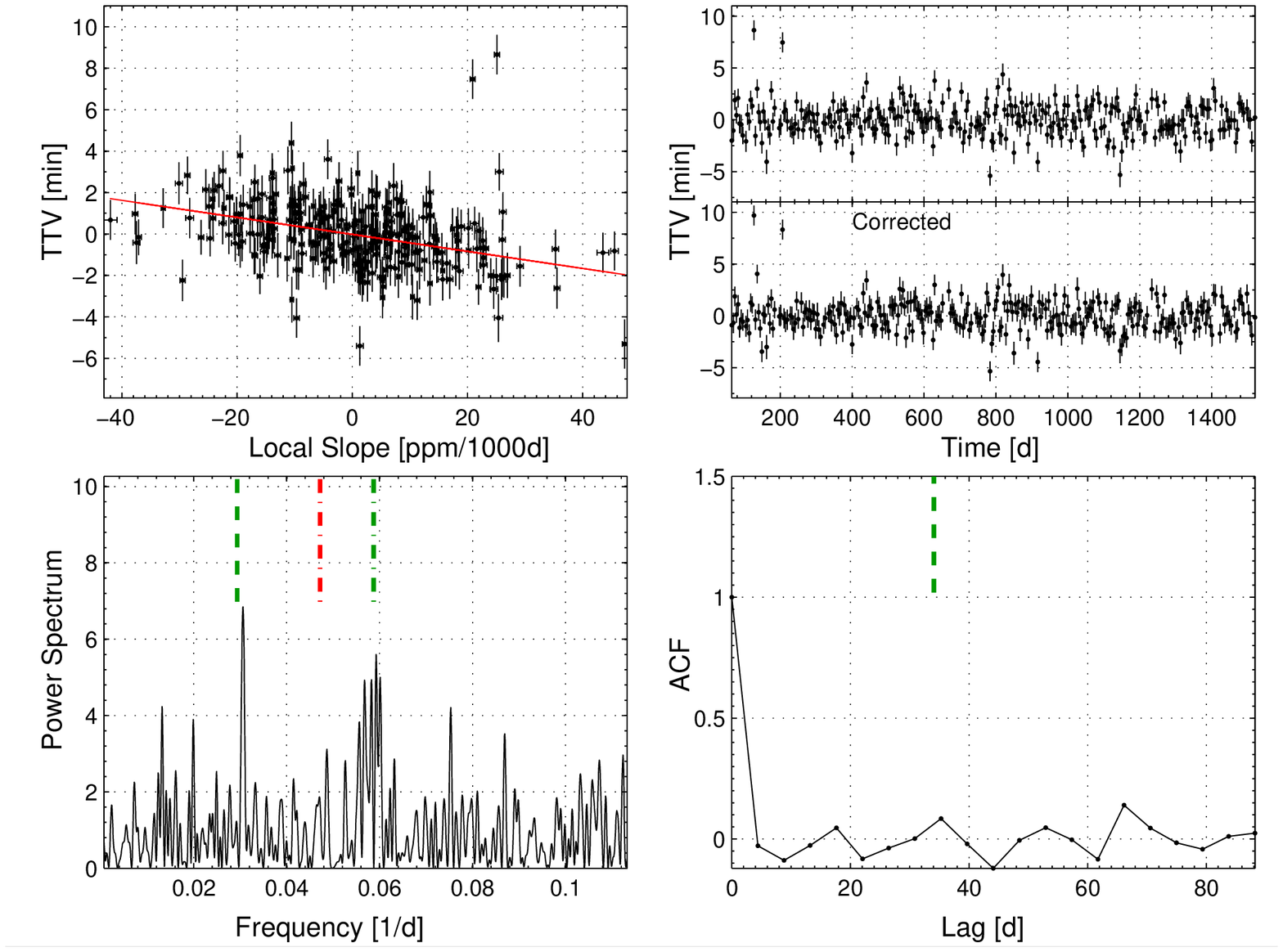}}
\caption{Analysis of the TTV of KOI-895.01. See \figr{koi1_2} for
details.
The green dashed line in the bottom left panel mark the stellar rotation frequency, and its first harmonic is marked by a green dotted-dashed line.
}
\label{fig:koi4_2}
\end{figure}

\begin{figure}
{\includegraphics[width=16.5cm]{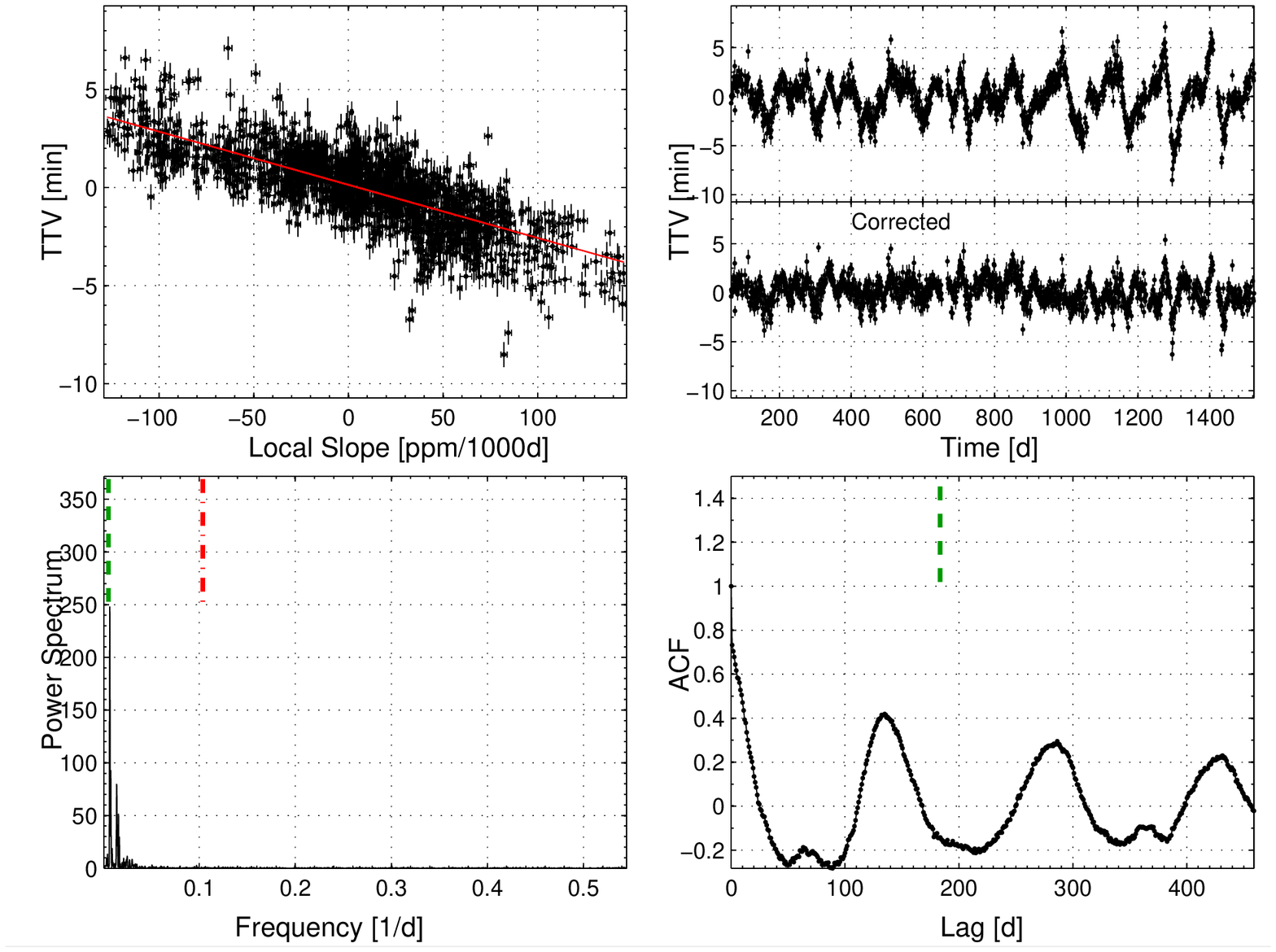}}
\caption{Analysis of the TTV of KOI-1546.01. See \figr{koi1_2}
for
details.
The green dashed line in the bottom left panel marks the stellar rotation frequency.
}
\label{fig:koi5_2}
\end{figure}

\begin{figure}
{\includegraphics[width=16.5cm]{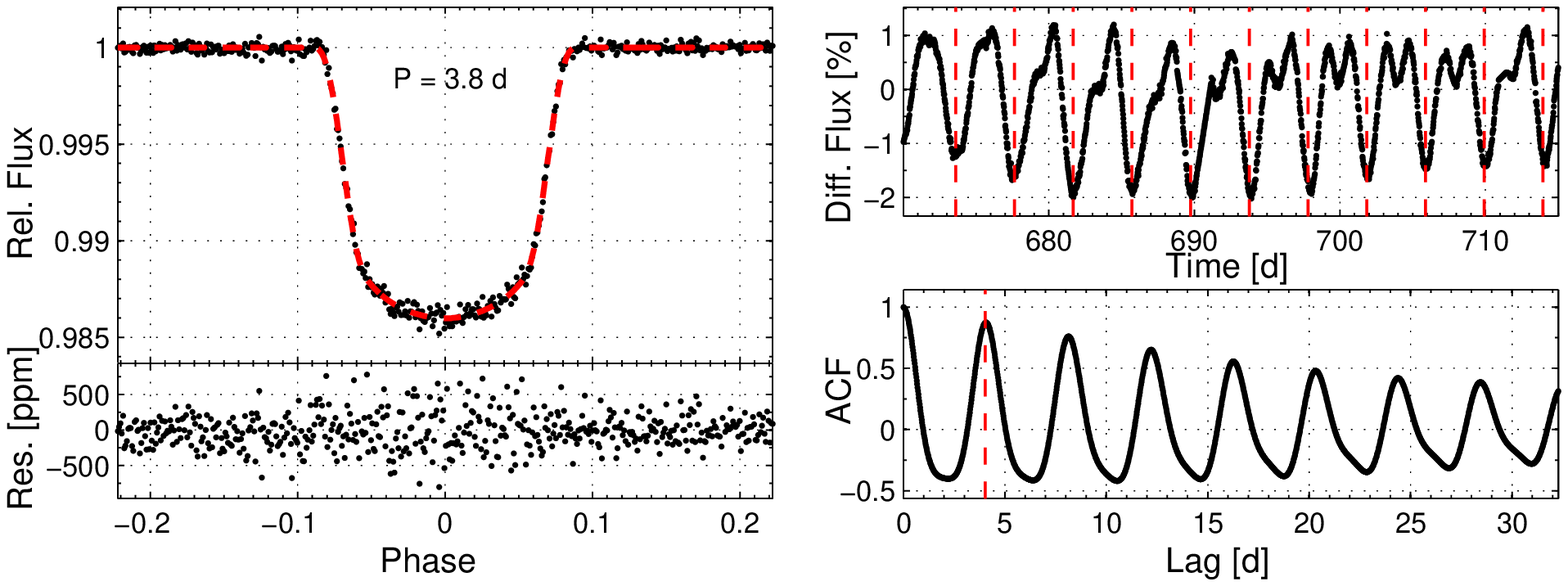}}
\caption{KOI-1074.01 transit light curve and stellar modulation
and its
autocorrelation. See \figr{koi1} for details.}
\label{fig:koi6}
\end{figure}

\begin{figure}
{\includegraphics[width=16.5cm]{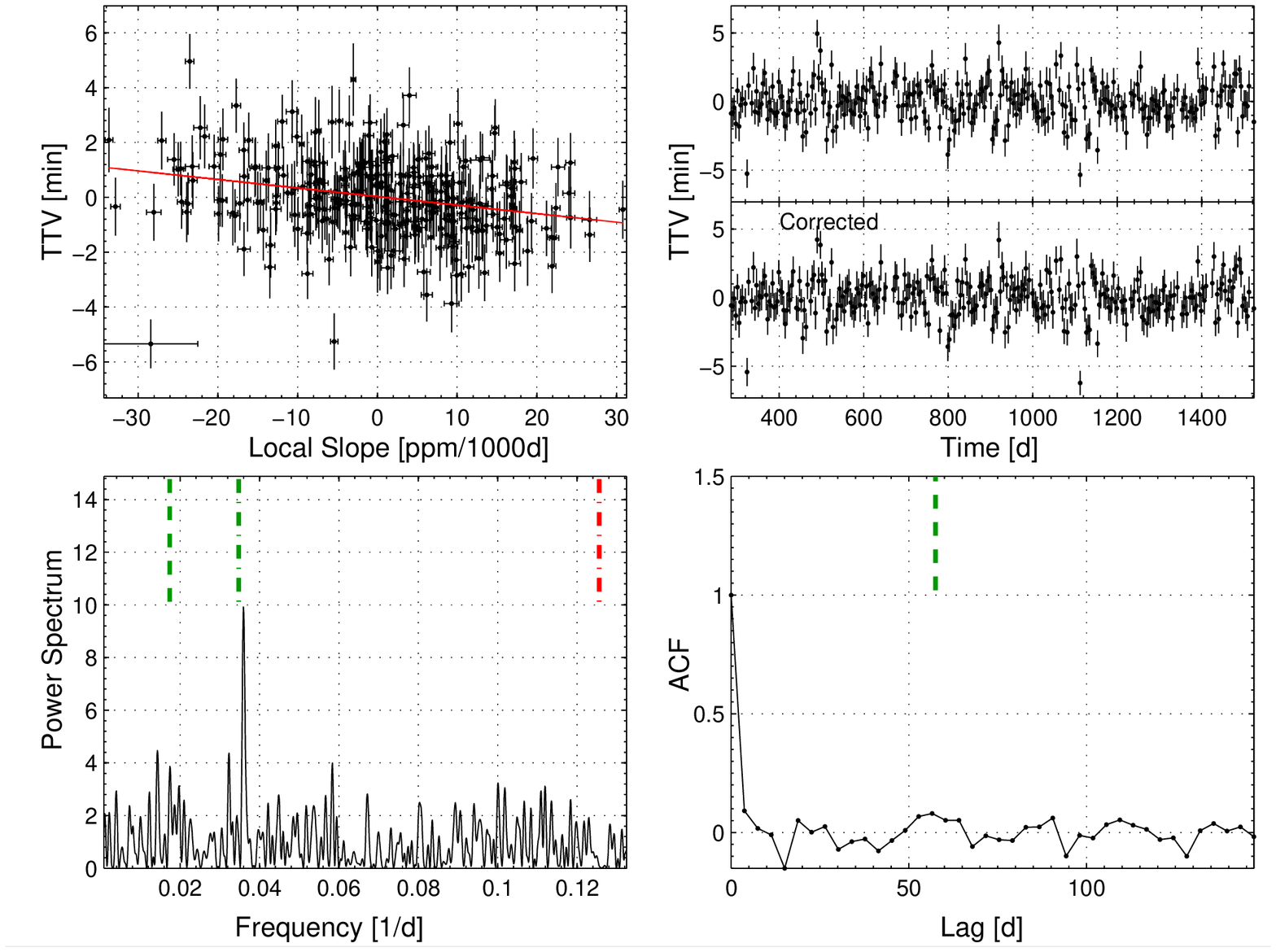}}
\caption{Analysis of the TTV of KOI-1074.01. See \figr{koi1_2}
for
details.
The green dashed line in the bottom left panel mark the stellar rotation frequency, and its first harmonic is marked by a green dotted-dashed line.
}
\label{fig:koi6_2}
\end{figure}

In this section we present the six systems that show
a highly significant anticorrelation between their TTV and the
corresponding local slope, with \gsr \pv lower than 10$^{-5}$.
Their analysis, based on the long cadence,  are presented in
Figures~\ref{fig:koi1}--\ref{fig:koi6_2}. Figures~\ref{fig:koi1},
\ref{fig:koi2}, \ref{fig:koi3}, \ref{fig:koi4},
\ref{fig:koi5} and \ref{fig:koi6} show the light curve analysis, and
Figures~\ref{fig:koi1_2}, \ref{fig:koi2_2}, \ref{fig:koi3_2},
\ref{fig:koi4_2}, \ref{fig:koi5_2} and \ref{fig:koi6_2} show the TTV analysis.

In Figures~\ref{fig:koi1}, \ref{fig:koi2}, \ref{fig:koi3},
\ref{fig:koi4}, \ref{fig:koi5} and \ref{fig:koi6}, the left panel presents the
phase folded light curve around the transit, overplotted by the fitted model
(red dashed line), with the residuals at the bottom. In all six systems, the larger
scatter in the residuals during transit compared to out of
transit is probably due to spot-crossing events. 

The top right panel presents a segment of the light curve as a function of time, where the photometric modulation due to stellar rotation
and activity is clearly noticeable. 
We can see that the modulation is not strictly periodic, as the spots evolve in size and location.
The spacing between the
vertical red dashed lines is the rotation period identified by
\citet{mcquillan13, mcquillan14}. 
They used the Auto-Correlation
Function (ACF), presented in the bottom right panel, where the red
dashed line marks the rotation period, which, as expected,
coincides with the shortest time lag ACF peak.

In Figures~\ref{fig:koi1_2}, \ref{fig:koi2_2}, \ref{fig:koi3_2},
\ref{fig:koi4_2}, \ref{fig:koi5_2} and  \ref{fig:koi6_2}, the top left panel shows
the TTV vs.~the local photometric slope with the linear fit
overplotted as a solid red line. The clear anti-correlation
between the TTV and the local slope indicates prograde motion.
The top right panel shows the measured TTV as a function of time
(top) and the corrected TTV (bottom) after subtracting the linear
fit of the TTV vs.~local slope. The corrected TTV does not show
much difference relative to the measured TTV (except maybe in
KOI-1546.01). 

The bottom left panel shows the TTV frequency power
spectrum, with a green dashed line marking the stellar rotation
frequency.  When relevant, another green dotted-dashed line or two mark the harmonics of this frequency.
The red dotted-dashed line marks the sampling
stroboscopic frequency---the TTV signal induced by the sampling
of the \ik long cadence data \citep{szabo13, mazeh13}. 
The power spectrum is plotted up to the Nyquist frequency, which in this case is half the orbital frequency, at which the TTV signal is sampled.
The bottom
right panel shows the TTV ACF with the stellar rotation period
marked by a green dashed line. 

In two cases, KOI-203 and KOI-883, we see a clear TTV periodicity with the rotational period. This is another indication that the TTV signal is related to the stellar rotation. 
In the case of KOI-217.01 and KOI-1074 (Figures~\ref{fig:koi2_2} and \ref{fig:koi6_2})  the strongest power spectrum (PS) peak is at twice the stellar rotation
frequency. This is consistent with a scenario in which the star has two significant
spots (or groups of spots), each at opposite hemispheres, as can be seen also in the stellar light curves (Figures~\ref{fig:koi2} and \ref{fig:koi6} top right panel). 

Note that for KOI-895, KOI-1074, and KOI-1546.01,  the rotational frequency $f_{\rm rot}$ is larger than the Nyquist frequency of the sampling, which is half the orbital frequency, $f_{\rm Nyq} = 0.5 f_{\rm orb}$ (see the orbital and rotation periods listed in \tabr{significant_TTV}). Therefore the rotational frequency is aliased to a smaller frequency. 

Out of the six KOIs, two are confirmed planet hosts.
KOI-$203.01$ (\figr{koi1}) was confirmed by
\citet[Kepler-17b]{desert11}, who used \ik short cadence data and
found it to be prograde with a spin-orbit angle smaller than 15$^{\circ}$, which agrees well with our findings. The second confirmed planet is KOI-217.01 \citep[Kepler-71b]{howell10}.

\section{KOIs with no significant correlation}
\label{sec:KOIs_non-significant}

\begin{figure}
{\includegraphics[width=16.5cm]{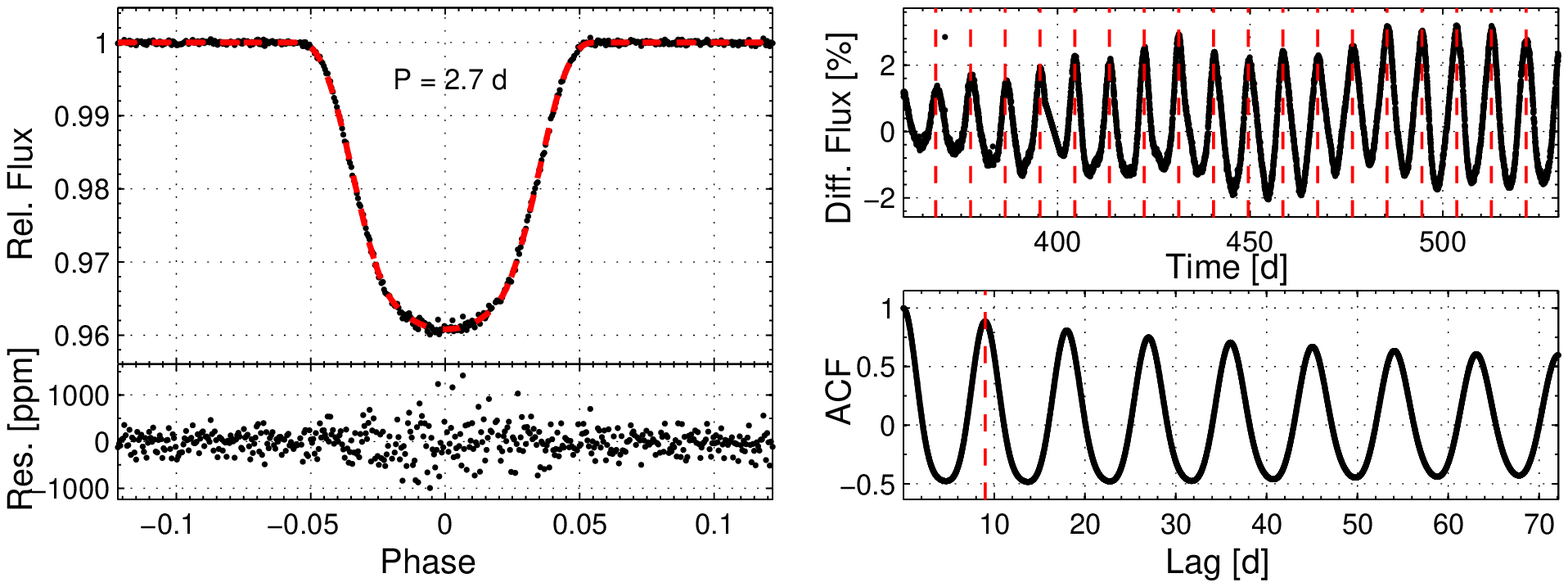}}
\caption{KOI-883.01 transit light curve and stellar modulation
and its
autocorrelation. See \figr{koi1} for details.}
\label{fig:koi3}
\end{figure}

\begin{figure}
{\includegraphics[width=16.5cm]{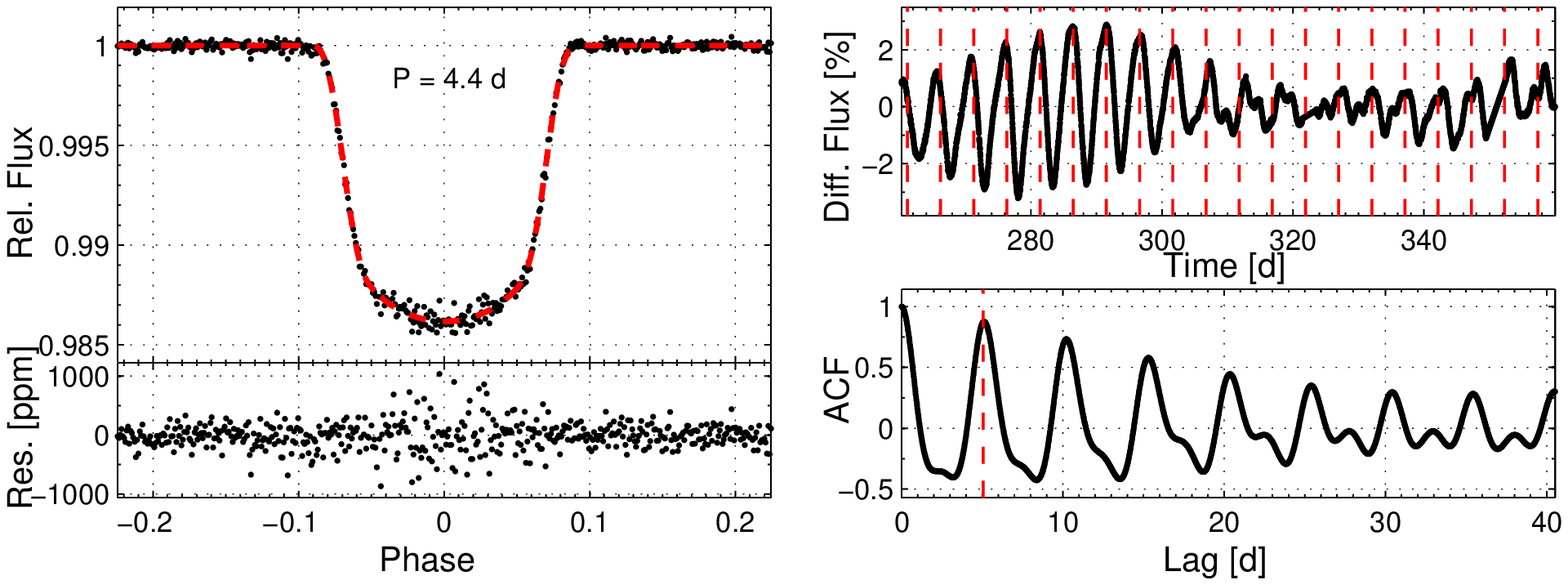}}
\caption{KOI-895.01 transit light curve and stellar modulation
and its
autocorrelation. See \figr{koi1} for details.}
\label{fig:koi4}
\end{figure}

\begin{figure}
{\includegraphics[width=16.5cm]{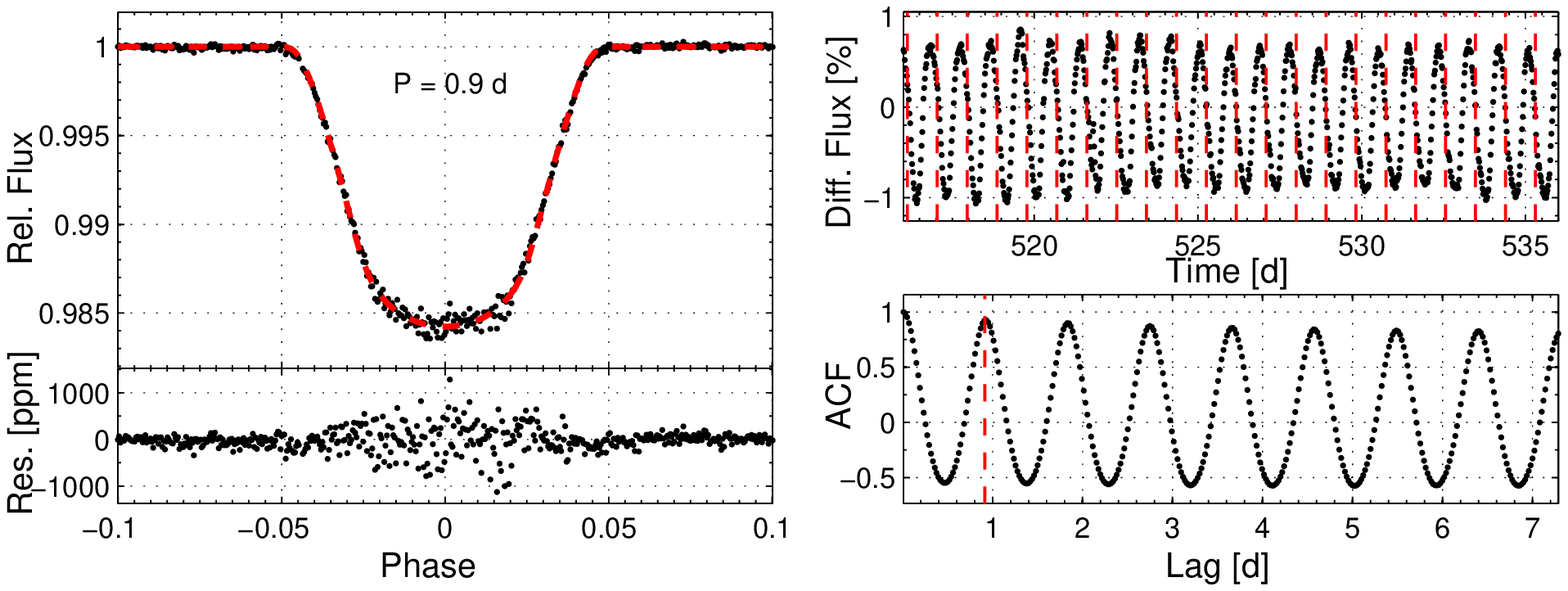}}
\caption{KOI-1546.01 transit light curve and stellar modulation
and its
autocorrelation. See \figr{koi1} for details.}
\label{fig:koi5}
\end{figure}

\begin{figure}
{\includegraphics[width=16.5cm]{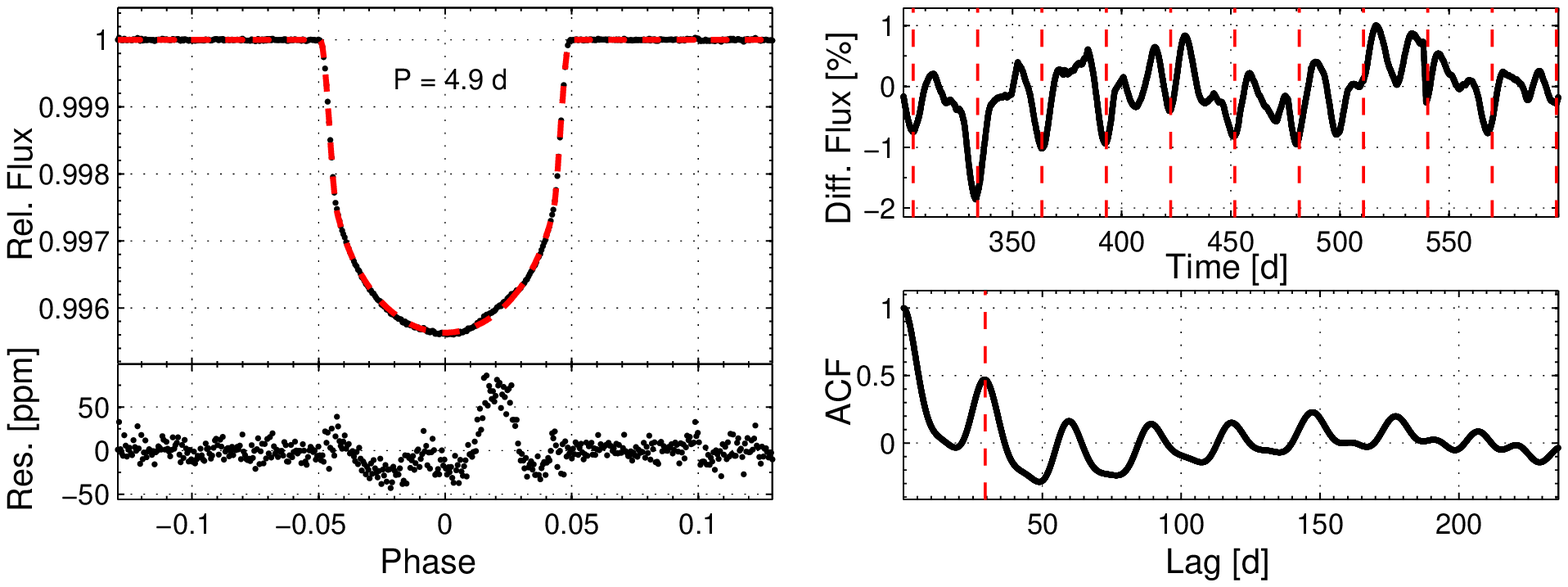}}
\caption{KOI-3.01 transit light curve and stellar modulation
and its
autocorrelation. See \figr{koi1} for details.}
\label{fig:koi7}
\end{figure}

\begin{figure}
{\includegraphics[width=16.5cm]{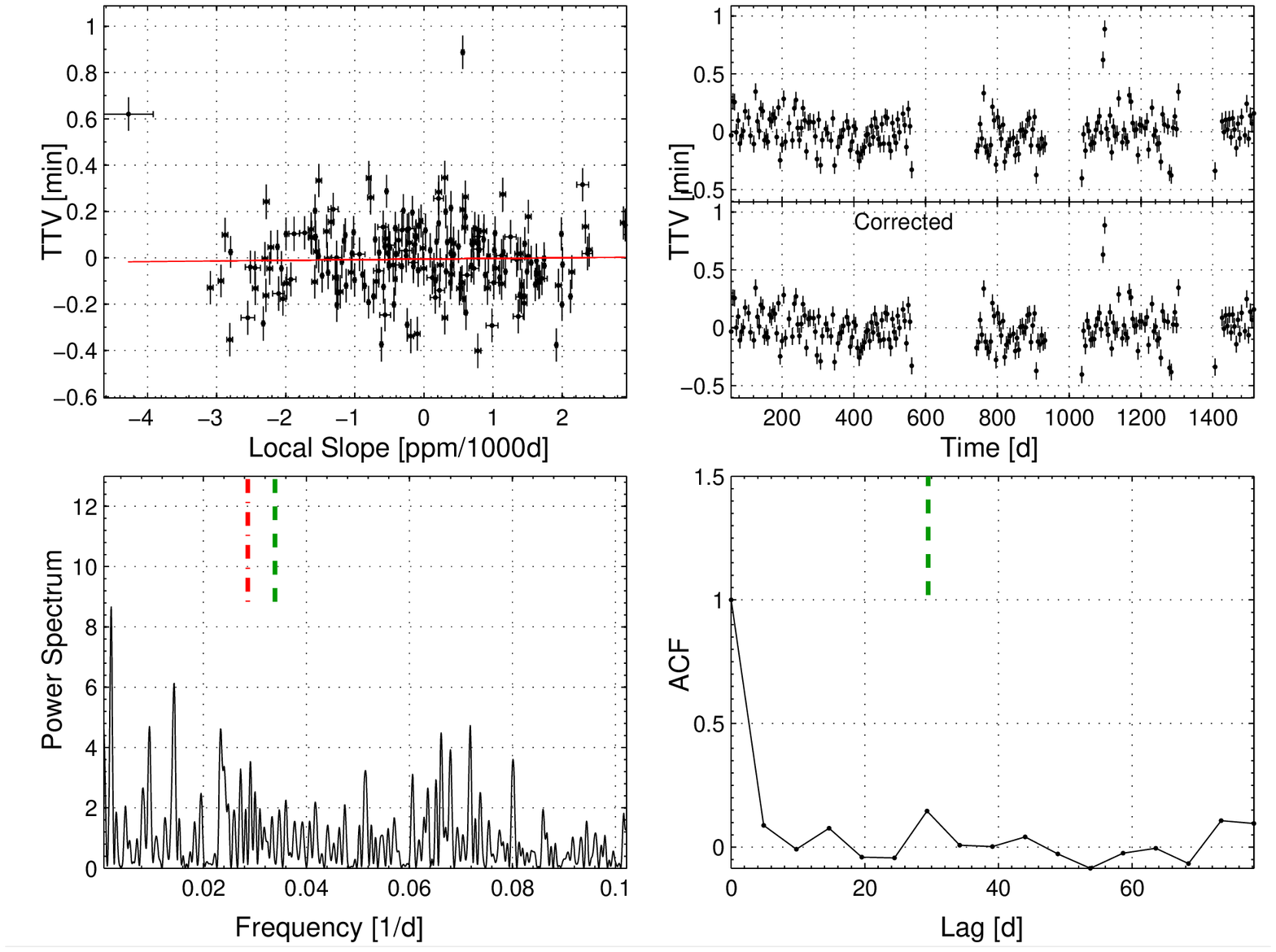}}
\caption{Analysis of the TTV of KOI-3.01. See \figr{koi1_2}
for
details.
The green dashed line in the bottom left panel mark the stellar rotation frequency.
}
\label{fig:koi7_2}
\end{figure}

\begin{figure}
{\includegraphics[width=16.5cm]{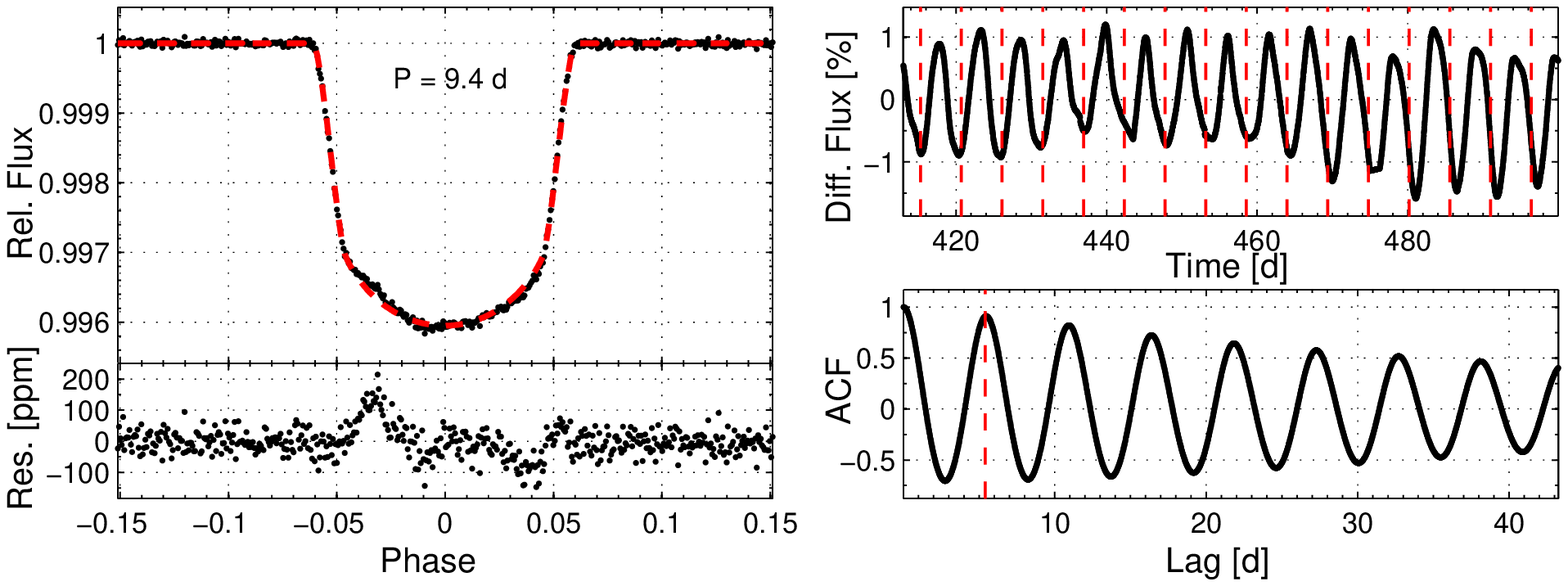}}
\caption{KOI-63.01 transit light curve and stellar modulation
and its
autocorrelation. See \figr{koi1} for details.}
\label{fig:koi8}
\end{figure}

\begin{figure}
{\includegraphics[width=16.5cm]{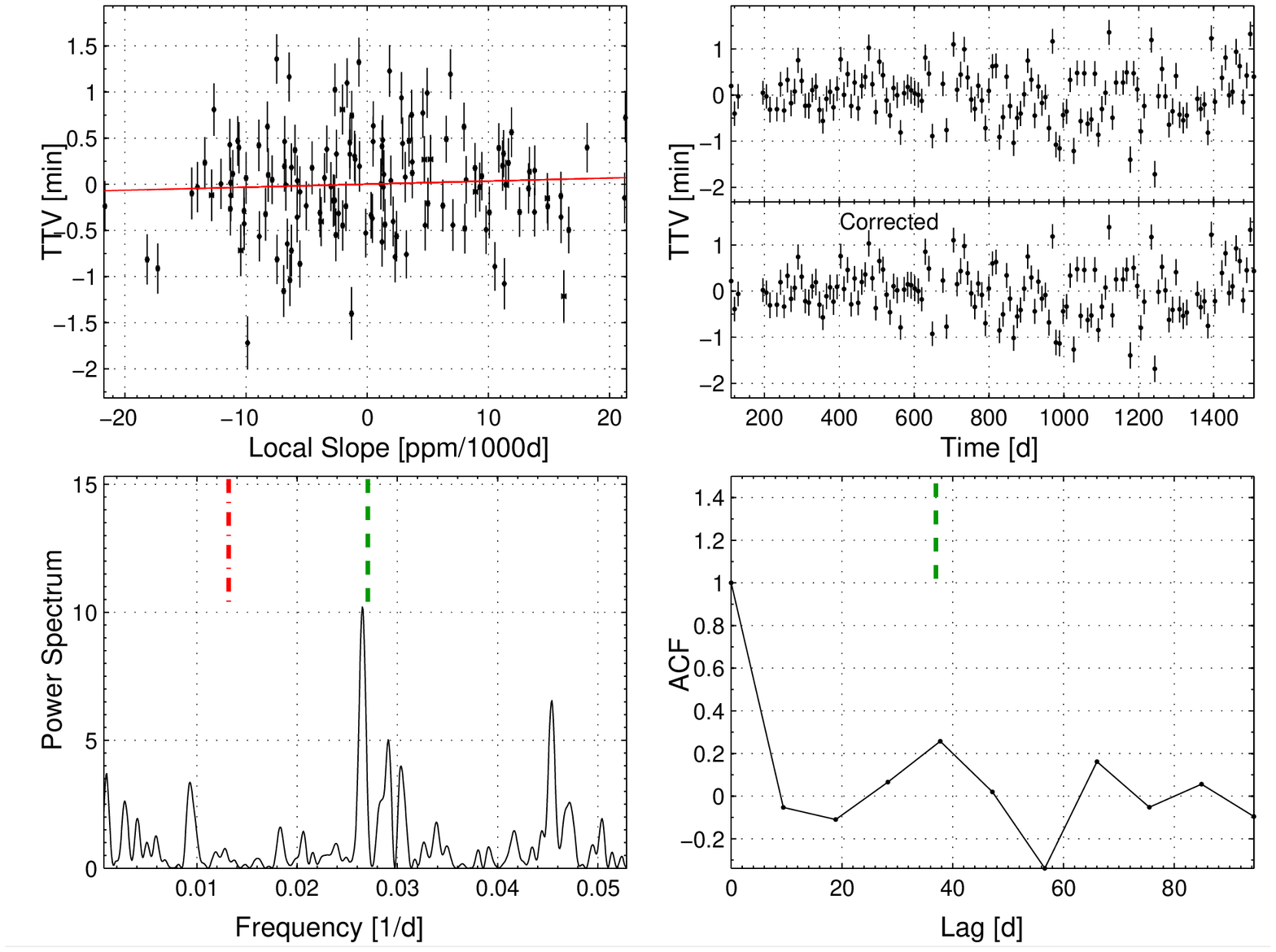}}
\caption{Analysis of the TTV of KOI-63.01. See \figr{koi1_2}
for
details.
The green dashed line in the bottom left panel mark the stellar rotation frequency.
}
\label{fig:koi8_2}
\end{figure}

\begin{figure}
{\includegraphics[width=16.5cm]{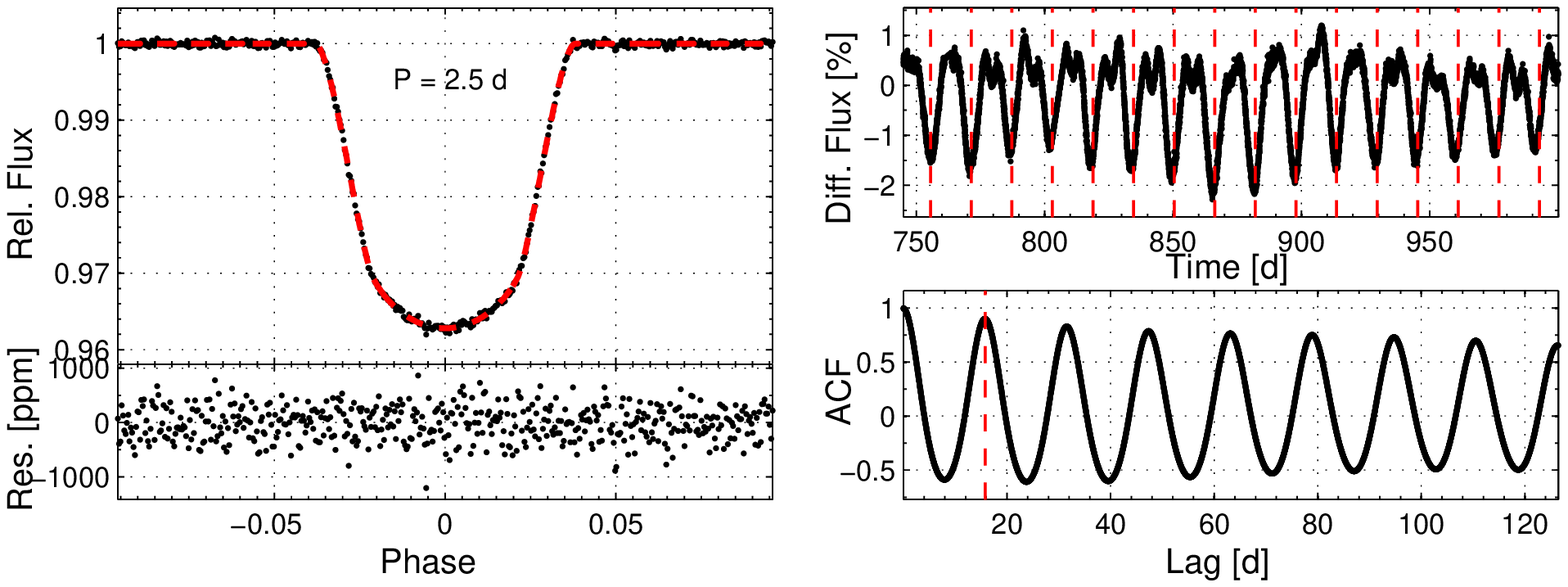}}
\caption{KOI-254.01 transit light curve and stellar modulation
and its
autocorrelation. See \figr{koi1} for details.}
\label{fig:koi9}
\end{figure}

\begin{figure}
{\includegraphics[width=16.5cm]{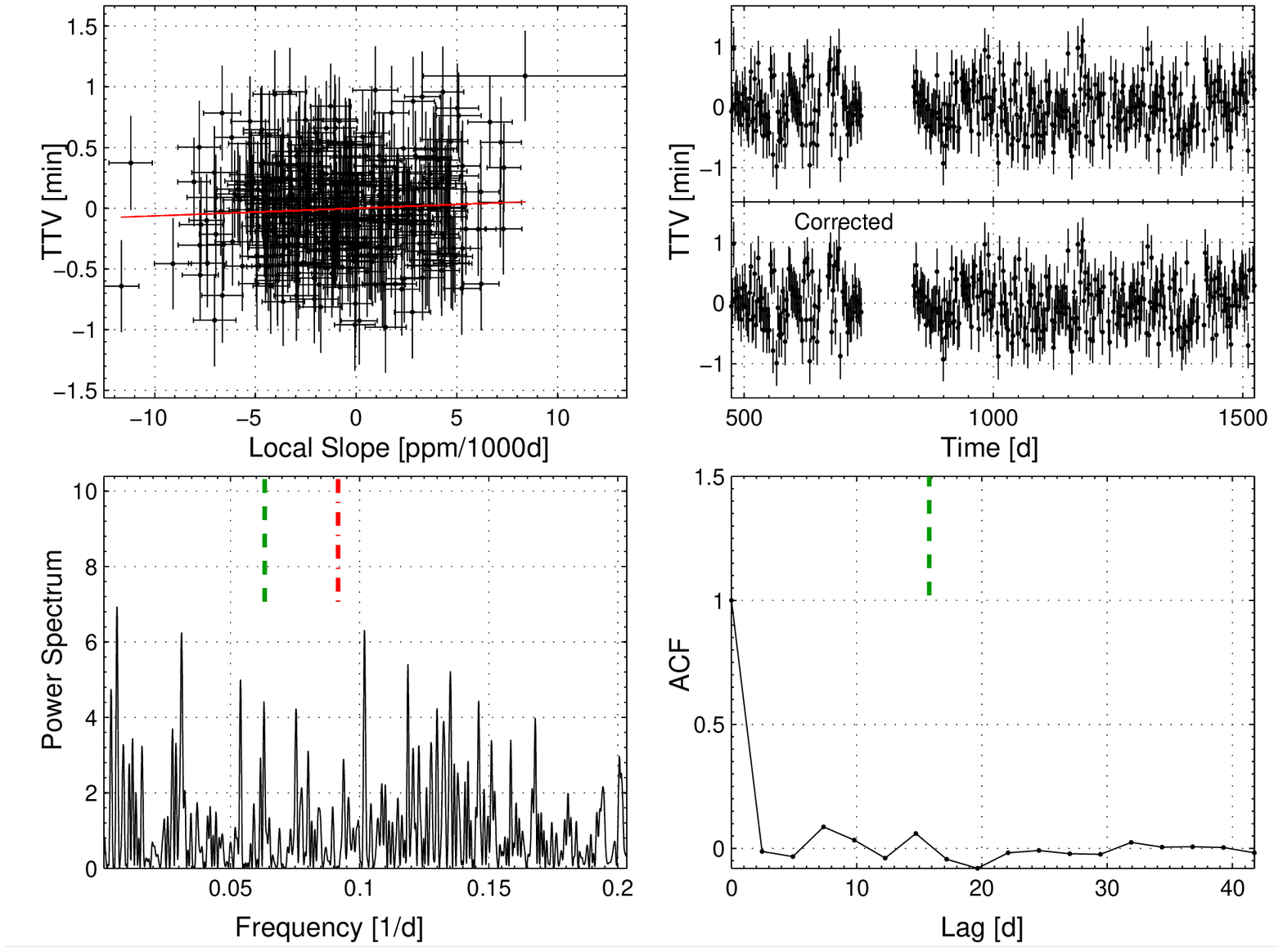}}
\caption{Analysis of the TTV of KOI-254.01. See \figr{koi1_2} for details. The green dashed line in the bottom left panel mark the stellar rotation frequency. 
}
\label{fig:koi9_2}
\end{figure}

In this section we present the three KOIs with high sensitivity but with no detected correlation between the TTV and the corresponding local-slope. All three,  KOI-3.01 (= Kepler-3b = HAT-P-11b; \citealt{bakos10}), KOI-63.01 (= Kepler-63b; \citealt{sanchis13}), and KOI-254.01 (= Kepler-45b; \citealt{johnson12}) are confirmed planets. They are presented in Figures~\ref{fig:koi7} through \ref{fig:koi9_2} in the same format as the KOIs showing significant correlation (see \secr{KOIs_significant} and Figures~\ref{fig:koi1} through \ref{fig:koi6_2}).  It can be visually seen that the TTV vs.~local slope does not show a significant correlation for all three systems (see Figures \ref{fig:koi7_2}, \ref{fig:koi8_2}, and \ref{fig:koi9_2}). 

In order to allow a more careful examination of the light curves of these three systems, and specifically to try and understand why they do not show a significant correlation, we use here the \ik short cadence data and the \cite{mandel02} transit model for generating Figures~\ref{fig:koi7} through \ref{fig:koi9_2}.

For KOI-3.01 (= Kepler-3b = HAT-P-11b; see Figures~\ref{fig:koi7} and \ref{fig:koi7_2}) the residuals of the phase folded and binned light curve are not consistent with white noise but instead show clear systematic features (\figr{koi7} left panel). This is in contrast to the other systems (see Figures \ref{fig:koi1}, \ref{fig:koi2}, \ref{fig:koi3}, \ref{fig:koi4}, \ref{fig:koi5}, and \ref{fig:koi6}) where the residuals of the in-transit light curve show an increased noise level but no systematic features. This distribution of the residuals suggests that the same features occur in many transit events at the same phases, so they do not average out in the folded light curve. Assuming these residuals are caused by spot-crossing events, this might suggest that the spot crossings occur primarily at specific phases during the transit. 

Fortunately, the \ik light curve of KOI-3.01 was carefully studied by several authors \citep{sanchis11b, deming11, beky14}, in addition to observations of the RM effect \citep{winn10b, hirano11}. These studies found that the system has  a sky-projected spin-orbit angle close to 90$^{\circ}$, and that the host star has two active latitudes. Hence, spot-crossing events occur when the planet moves across the active stellar latitudes, explaining why they are seen only at specific phases during the transit. Since those phases sample only part of the transit and not the entire transit it reduces the TTV range induced by the spot-crossing events. This is also consistent with a weak TTV periodicity identified at the rotation period (see \figr{koi7_2}), where no significant peak is seen at the rotation frequency (bottom left panel) and only a small peak, close to the noise level at the ACF (bottom right panel). Moreover, in this configuration, when a spot is crossed by the planet the star is at a similar phase in its rotation, resulting in a small range of local slopes. The combined result is that although this system posses the sensitivity for TTV--local-slope correlation, its characteristics suppress that correlation.

KOI-63.01 (= Kepler-63b; see Figures~\ref{fig:koi8} and \ref{fig:koi8_2}) also shows systematic features in the phase folded light curve residuals (\figr{koi8} left panel). \cite{sanchis13} have already studied the Kepler-63b \ik light curve and observed the RM effect. They determined the system has a sky-projected spin-orbit misalignment of -110$^{+22}_{-14}$ deg and a stellar spin axis inclination angle of 138$\pm$7 deg. Therefore, in this configuration the spot-crossing events are confined to the same parts of the transit. This explains the features in the phased light curve residuals and the non-detection of a spots-induced TTV signal, where the latter leads to no TTV--local-slope correlation. Yet, the TTV period analysis (\figr{koi8_2}) does show a significant TTV periodicity at the rotation frequency, which in this case is aliased to $f_{\rm rot, alias} = 2f_{\rm orb} - f_{\rm rot}$,  as 
$3f_{\rm Nyq}<f_{\rm rot} <4f_{\rm Nyq}$. This could be explained by assuming there are one or two spot latitudes and therefore one or two transit phases that allow spot crossing, not enough to produce a detectable correlation, but enough to induce a TTV modulation with the rotation period.

For KOI-254.01 (= Kepler-45b, \citealt{johnson12}; see Figures~\ref{fig:koi9} and \ref{fig:koi9_2}), the spin-orbit angle is currently not known. We hypothesize that the planet does not cross spots during the transit and that the system is spin-orbit aligned, similar to, e.g., Kepler-77 \citep{gandolfi13}. This is consistent with the noise level of the in-transit light curve being similar to the noise level of the out-of-transit phases, which is different from the six systems with significant correlation that show an increased in-transit residuals noise level (see \secr{KOIs_significant} and Figures \ref{fig:koi1}, \ref{fig:koi2}, \ref{fig:koi3}, \ref{fig:koi4}, \ref{fig:koi5}, and \ref{fig:koi6}). For this system we have determined here a transit impact parameter of 0.5732 $\pm$ 0.0027, consistent with \citealt{johnson12} value of 0.6 $\pm$ 0.2, who used less \ik data than used here. This impact parameter is larger than the ones of the systems with significant correlation (see \figr{S_vs_b}), suggesting that KOI-254.01 impact parameter is larger than the impact parameter corresponding to the typical active stellar latitudes. For this object the TTV period analysis (\figr{koi9_2}) does not show a periodicity at the rotation period, as expected.

Another possible scenario that can weaken the correlation between the TTV and the local slope and was not mentioned above is a dynamical TTV signal, following the gravitational interaction with another body in the systems \citep[e.g.,][]{fabrycky12, ford12, steffen12}. However, there is no evidence for such a signal in any of the three systems discussed above. 

To summarize this section, for all three systems we are able to provide plausible scenarios that explain why despite having the sensitivity to a TTV signal induced by spot-crossing events, these systems do not show a detected correlation.

\section{Discussion}
\label{discussion}

We presented here the application of a simple method to distinguish between
prograde and retrograde transiting star-planet systems, 
where the brightness of the host star is
modulated by stellar spots, and the transits include spot-crossing events. The method is based on the assumption that even when
spot-crossing events are not resolved,
they induce a shift in the derived mid-transit timing, when
fitting a model that ignores the spot-crossings. We have applied our
method to a sample of 2,600 KOIs whose transit timings were
measured by \cite{holczer15} using the long cadence data, and are publicly available
(ftp://wise-ftp.tau.ac.il/pub/tauttv/TTV/ver$\_$112). 
We have concentrated on 862 systems with 
published stellar rotation periods \citep{mcquillan13,mcquillan14}, out of which only 726 had at least 20 measurements of the TTV and local slope.

Using the formalism of Paper I, we identified nine systems,
listed in Table 3,
with high enough stellar modulations and large enough planets to allow the detection of the TTV local-slope correlation. 
The nine systems have relatively large photometric modulation amplitudes due to rotation
($\gtrsim 5000$ ppm), all of them
well above the median of the whole analyzed sample ($\sim$3000 ppm).
 The precision of the transit timing
for these nine systems is high, with a median of
0.45 min, while the sample median is $\sim$13 min. 
This indicates that our method requires objects showing relatively strong rotation modulation and
high SNR transits, the latter leading to highly precise
mid-transit timings. The nine systems we have identified have orbital periods in the range of 0.9--9.4~d and transit depth of 0.4--4.0\%, consistent with the typical parameters range of hot Jupiters. The small size of the sample is not surprising, given the fact that we are looking for a minor effect that is hiding in the noise in most cases. One of the nine cases is KOI-1546, which recently has been found to be a stellar binary, so we are left with eight {\it bona fide} planet-candidate systems (five of which already confirmed as planets).

It turned out that out of these eight systems five show a clear negative correlation, indicating prograde motion, and none a positive correlation, which could have indicated a retrograde motion. This is consistent with the finding of the seminal work of \citet{winn10a}, who showed that most cool planet-host stars have close to zero obliquity, as four out of our five systems have lower surface temperature. 
However, this could also result from an intrinsic bias in our approach, since it requires active host stars, which are typically convective and hence relatively cold.

As shown in Figure~\ref{fig:S_vs_b}, the five systems with detected correlation (not including KOI-1546, which is a binary) 
{\it all} have an impact parameter in the range of $0.2\lesssim b\lesssim 0.5$. Assuming those systems have their spots moving on similar latitudes as the planetary chord as seen from Earth, this result shows a similarity to the Sun, where the spots reach up to about $30^\circ$ latitude and decrease in the size as they move closer to the Sun's equator.

To summarize, we have found evidence for prograde motion of transiting planets  but no retrograde motion. Furthermore, we found some hints that stellar spots might preferentially be at 
low, Sun-like, latitudes on the stellar disc ($<30^\circ$) but not exactly along the stellar equator, although our search suffers from an observational bias against
detection of high latitude spots, and the detected sample is extremely small. 
The original goal of this project was to learn about the planetary motion, but we were surprised to find a possible hint about a feature of spot motion. In principle this is possible, as the planet with its small radius is sampling the stellar surface with high resolution that is not available in present-day observational techniques. The high precision, evenly spaced, long-span light curves of \ik make this possible.  

The approach described here is not limited to \ik star-planet
systems and can, in principle, be applied to \ik stellar binary
systems \citep{prsa11, slawson11}, an analysis that we are
preparing in a forthcoming paper. In fact, a negative correlation between TTV and local
slope was identified already for the eclipsing stellar binary
system within the
circumbinary planetary system Kepler-47 \citep{orosz12}, where a
more detailed analysis of the spot-crossing events indicates a prograde motion
\citep{orosz12}.

In the future the method used here can be applied to a large
sample of systems monitored by
current and future space-based surveys delivering high-quality
photometry, including K2 \citep{howell14}, TESS \citep{ricker14},
and PLATO \citep{rauer14}. A larger sample can reveal the typical
characteristics of prograde and retrograde systems, which in turn
will allow constraining formation and orbital evolution
processes. 
In addition, a larger sample of systems where spot-crossing events are identified will also better constrain spots' behavior.

\acknowledgments
We wish to warmly thank Jason Rowe, Fergal Mullally, and Jack Lissauer, for discussions and feedback that helped improve this paper.
The research leading to these results has
received funding from the European Research Council under the
EU's Seventh Framework Programme (FP7/(2007-2013)/ ERC Grant
Agreement No.~291352).
T.M. acknowledges support from the Israel Science
Foundation (grant No.~1423/11) and the Israeli Centers of
Research Excellence (I-CORE,
grant No.~1829/12).
E.B.F. was supported in part by NASA \ik Participating Scientist Program award NNX12AF73G, NASA Origins of Solar Systems award NNX14AI76G, and NASA Exoplanet Research Program award NNX15AE21G.  
The Center for Exoplanets and Habitable Worlds is supported by the Pennsylvania State University, the Eberly College of Science, and the Pennsylvania Space Grant Consortium.
W.F.W and J.A.O gratefully acknowledge support from the NSF via grant AST-1109928,
and from NASA via grants NNX13AI76G-3 and NNX14AB91G.
D.F. was supported by the National Aeronautics and Space Administration under Grant No. NNX14AB87G issued through the Kepler Participating Scientists Program. 
The last phase of this study was done when T.M. and R.S. were members of the KITP program of
``Dynamics and Evolution of Earth-like Planets". They wish to thank the director of KITP, Lars Bidsten, and the coordinators of the program,  Eric Ford, Louise Kellogg, Geoff Marcy, and Burkhard Militzer, for participation in the program.
This work was performed in part
at the Jet Propulsion Laboratory, under contract with
the California Institute of Technology (Caltech) funded
by NASA through the Sagan Fellowship Program executed
by the NASA Exoplanet Science Institute.
All photometric data presented in this paper were
obtained from the Mikulsky Archive for Space Telescopes (MAST).
STScI is operated by the Association of Universities for Research
in Astronomy, Inc., under NASA contract NAS5-26555. Support for
MAST for non-HST data is provided by the NASA Office of Space
Science via grant NNX09AF08G and by other grants and contracts.





\end{document}